\documentstyle[preprint,tighten,aps]{revtex}

\begin{document}
\draft
\preprint{IFUP-TH 52/97}
\title{
Four-point renormalized coupling constant 
and Callan-Symanzik $\beta$-function in O($N$) models.
}
\author{Andrea Pelissetto and Ettore Vicari}
\address{Dipartimento di Fisica dell'Universit\`a 
and I.N.F.N., I-56126 Pisa, Italy}

\date{\today}

\maketitle

\begin{abstract}
We investigate some issues concerning the zero-momentum
four-point renormalized coupling constant $g$ in 
the symmetric phase of O$(N)$ models,
and the corresponding Callan-Symanzik $\beta$-function. 

In the framework of the $1/N$ expansion we show 
that the Callan-Symanzik $\beta$-function
is non-analytic at its zero, i.e. at the fixed-point
value $g^*$ of $g$.

This fact calls for a check of the actual accuracy of
the determination of $g^*$ from 
the resummation of the $d=3$ perturbative $g$-expansion,
which is usually performed assuming the analyticity of the
$\beta$-function. Two alternative approaches are exploited.

We extend the $\epsilon$-expansion of $g^*$ to $O(\epsilon^4)$.
Quite accurate estimates of $g^*$ are obtained 
by an analysis that exploits the analytic behavior of $g^*$
as a function of $d$ and the 
known values of $g^*$ 
for lower-dimensional O$(N)$ models, i.e. for $d=2,1,0$. 

Accurate estimates of $g^*$ are also obtained by a
reanalysis of the strong-coupling expansion
of the lattice $N$-vector model
allowing for the leading confluent singularity.

The agreement among the $g$-,
$\epsilon$-, and strong-coupling expansion results is good
for all values of $N$. However,
at $N=0,1$, $\epsilon$- and strong-coupling expansion
favor values of $g^*$ which are slightly lower than those
obtained by the resummation of the $g$-expansion assuming
analyticity in the Callan-Symanzik $\beta$-function.


\end{abstract}

\newpage
\newcommand{\N}{\hbox{{\rm I}\kern-.2em\hbox{\rm N}}}

\section{Introduction}
\label{introduction}

The renormalization-group theory of critical phenomena
provides a description of statistical models 
in the neighbourhood of the critical point.
For ${\rm O}(N)$ models calculations are based on
the $\phi^4$-field theory defined by the action
\begin{equation}
S= \int d^dx\left[ {1\over 2}\partial_\mu \phi(x)\partial_\mu \phi(x)
+ {1\over 2}m_0^2\phi^2 + {1\over 4!}g_0(\phi^2)^2\right].
\label{Sphi4}
\end{equation}
A strategy, which has been largely employed in the
study of the symmetric phase, relies on a perturbative expansion
in powers of the 
zero-momentum four-point renormalized coupling constant $g$
performed at fixed dimension $d=3$~\cite{Parisig}.
This perturbative expansion is asymptotic; 
nonetheless accurate results can be obtained
by resummations exploiting its Borel summability and the knowledge
of the large-order behavior.
As general references on the
$g$-expansion  method see for instance
Refs.~\cite{ZJbook,Parisibook,Itz-Dr}.
This technique has led to accurate estimates
of the critical exponents.

An important quantity entering the calculation of universal quantities
is the fixed-point value of $g$, i.e. the zero
of the corresponding Callan-Symanzik $\beta$-function. 
In the critical region, the bare coupling constant $g_0$
becomes infinite on the scale fixed by the correlation length,
whereas the zero-momentum four-point renormalized coupling
approaches a finite non-zero limit $g^*$ at criticality. 
Accurate calculations of $g^*$ 
have been done by analyzing the perturbative expansion  
of the Callan-Symanzik 
$\beta$-function~\cite{Nickel,LeG-ZJ,Nickel91,Nickelunp,Antonenko,Sokolov}
(known to $O(g^7)$),
using the known results for its large-order behavior. 
The best determinations of $g^*$ have been apparently obtained
by Le Guillou and Zinn-Justin~\cite{LeG-ZJ} by making some additional
assumptions on the analytic properties of the
Borel transform. Such additional
assumptions have been questioned by Nickel~\cite{NickelCS,Nickel91},
who argued the presence of confluent singularities
in the $\beta$-function at its zero,
which may complicate the analytic structure of the 
Borel transform.

This issue needs a non-perturbative analysis to be clarified, furthermore
an analytic approach is required in order to understand the nature of the
singularities. These features can be realized in the framework of the
$1/N$ expansion. In this paper we
analyze the Callan-Symanzik $\beta$-function
computed in Ref.~\cite{ONgr} to $O(1/N)$ (i.e. the next-to-leading order). 
While the leading order is analytic, the $O(1/N)$ term
shows the presence of confluent singularities at the zero
of the $\beta$-function for all $2<d<4$.
Moreover a phenomenon analogous to the Abe-Hikami anomaly
for the specific-heat~\cite{Abe-Hikami} emerges at the special 
dimensions $d=4-2/n$ (where $n$ is an integer number), 
thus including the interesting case $d=3$.

In the analysis of Ref.~\cite{LeG-ZJ} confluent singularities
at the zero of the $\beta$-function may cause
a slow convergence to the correct fixed-point
value of $g$~\cite{NickelCS}. 
The apparent stability of the results when analyzing
a finite number of terms of the perturbative expansion may
then not provide a reliable indication of the uncertainty of the
overall estimate. 
Confluent singularities represent a source
of systematic error for the procedure used in Ref.~\cite{LeG-ZJ}.
A more general analysis explicitly allowing for the presence of
confluent singularities
would slightly change the value of $g^*$ 
for small values of $N$ (although not excluding the
values obtained in Ref.~\cite{LeG-ZJ}) and consequently
the values of the critical exponents~\cite{Nickelunp}.
It is therefore important
to exploit other approaches to the study of O($N$) models,
which can provide a check of the estimates of $g^*$ 
from the resummations of the perturbative
$g$-expansion, and of their actual accuracy.

An alternative field-theoretic
strategy is the expansion in powers of 
$\epsilon=4-d~\cite{Wilson}$.
An important
advantage of the $\epsilon$-expansion is the possibility
of working directly at criticality. 
This allows us to go from one phase to another using the same framework.
In order to get estimates at $\epsilon=1$, 
the $\epsilon$-expansion requires eventually a resummation
which is usually performed assuming its Borel summability. 
Relatively long series of the critical exponents have been calculated 
and their analysis has led to estimates which are in substantial
 agreement with those determined by the $g$-expansion
(see e.g. Ref.~\cite{eexp}).

The fixed-point value of the zero-momentum four-point 
renormalized coupling 
is known only to $O(\epsilon^2)$~\cite{B-L-Z-2}, thus not allowing a 
real check of the value of $g^*$ obtained by the $g$-expansion. 
In this paper we extend this calculation to $O(\epsilon^4)$.
The analysis of the $O(\epsilon^4)$ series of $g^*$
provides already good estimates with an apparent 
uncertainty of approximately 6\% for small values of $N$.
A considerable improvement in the analysis of the $\epsilon$-expansion
of $g^*$ is achieved using the known values of
$g^*$ for lower-dimensional $O(N)$ models, i.e. for $d=2,1,0$.
The key point is that $g^*$
is expected to be analytic and quite smooth in the domain
$0<d< 4$ (thus $0 < \epsilon < 4$). 
This can be indeed verified in the 
large-$N$ limit to $O(1/N)$ using the results of Ref.~\cite{ONgr}.
Generalizing the technique presented in Ref.~\cite{eexp},
we perform a polynomial interpolation among 
the values of $d$ where $g^*$ is known ($d=0,1$) or
for which good estimates are available ($d=2$, especially
for $N=0,1$ by strong-coupling calculations), 
and then analyze the series of the difference.
This procedure leads to much more accurate estimates of $g^*$,
which are consistent with those obtained by the direct analysis
of the original $\epsilon$-series, but with an apparent error of 
approximately one per cent ($\sim 0.5\%$ for $N=1$).
The agreement with the $g$-expansion estimates is 
good for all values of $N$.  However it is worth anticipating
that the results for $N=0,1$ turn out slightly
lower than the estimates given by Le Guillou and Zinn-Justin~\cite{LeG-ZJ},
thus favouring the results of 
the more general analysis done by Nickel~\cite{Nickelunp}.

Another approach which has been widely used in the study of critical
phenomena is based on lattice formulations of the theory.
Two main techniques have been exploited
in this context: high- and low-temperature
expansions and Monte Carlo simulations.
In the symmetric phase
the fixed-point value of the
zero-momentum four-point renormalized coupling 
can be estimated by
analyzing the strong-coupling expansion of the quantities 
entering its definition
\cite{Bakerold,Nickel-Sharpe,Bakergr2,B-etal,ONgr,Reisz,Z-L-F,B-C-g}.
Studies based on Monte Carlo simulations can be found in 
Refs.~\cite{Freedmanetal,Freedman-Baker,Wheater,Weston,K-P,Tsypin,BakerKawa,K-L}. 
The agreement with the field-theoretic estimates
is substantially good,
but small discrepancies have been  observed, especially for small values
of $N$~\cite{ONgr,Z-L-F,BakerKawa,K-L}. 
We will show that such residual discrepancies 
disappear (or are largely reduced) 
when the leading effects of the confluent singularities are properly
taken into account in the analysis of the strong-coupling 
expansion~\cite{Roskies,A-M-P,B-C-g}
and in the analysis of the Monte Carlo data. 

The paper is organized as follows:

In Sec.~\ref{sec2} we introduce some definitions and notations
used in the paper.

In Sec.~\ref{sec3} we analyze the Callan-Symanzik $\beta$-function
calculated to $O(1/N)$ in Ref.~\cite{ONgr}, and discuss the presence
of confluent singularities at its zero.

In Sec.~\ref{sec4} we analyze the expansion to 
$O(\epsilon^4)$ of the fixed-point value of the zero-momentum
renormalized coupling constant. 

Sec.~\ref{sec5} is dedicated to a reanalysis of the 
strong-coupling expansion which allows 
for the confluent singularities present
in the lattice approach. 

In Sec.~\ref{sec6} some conclusions are drawn.

In App.~\ref{appunosuN} we give some technical details 
on the analysis of the analiticity properties of the
Callan-Symanzik $\beta$-function to $O(1/N)$.

In App.~\ref{Appepsilon} 
we present the perturbative calculation to $O(\epsilon^4)$ of $g^*$.

\section{Definitions and notations}
\label{sec2}

Let us introduce a few definitions and notations. If we set
\begin{eqnarray}
&&\langle \phi^\alpha(0) \phi^\beta(x) \rangle
= \delta^{\alpha\beta} G^{(2)}(x) ,\\
&&\langle \phi^\alpha(0) \phi^\beta(x) \phi^\gamma(y) \phi^\delta(z) \rangle_c
= 
  {1\over3}(\delta^{\alpha\beta}\delta^{\gamma\delta} + 
            \delta^{\alpha\gamma}\delta^{\beta\delta} + 
            \delta^{\alpha\beta}\delta^{\gamma\delta} ) 
            G^{(4)}(x,y,z)  ,
\end{eqnarray}
the zero-momentum four-point renormalized coupling $g$ is defined as 
\begin{equation}
g  \equiv -  {\int dxdydz\ G^{(4)}(x,y,z) \over 
     \xi^d \left[ \int dx\ G^{(2)}(x) \right]^2}  ,
\label{gdef}
\end{equation}
where $\xi$ is the second-moment correlation length
\begin{equation}
\xi^2 = {1\over 2d} {\int dx \ x^2 G^{(2)}(x) \over \int dx\ G^{(2)}(x)} .
\end{equation}
The normalization in Eq.~(\ref{gdef}) is such that in perturbation theory
$g = g_0/m_0^\epsilon + O(g_0^2/m_0^{2\epsilon})$. 

For convenience in the paper we will also introduce other definitions.
First of all we consider the rescaled coupling
\begin{equation}
\bar{g} \equiv 
   {1\over 2 (4\pi)^{d/2}} {(N+8)\over3} \Gamma\left(2 - {d\over2}\right)g.
\label{gbar}
\end{equation}
Unlike $g^*$ which is of order $(4-d)=\epsilon$ for $d\to 4$,
the fixed-point value of $\bar{g}$ is $O(1)$, 
due to the factor multiplying $g$ in its definition. 
Moreover $\bar{g}$ has the property that for $N\to\infty$,
$\bar{g}^*\to 1$ for any dimension $d$.
For $d=3$,
it coincides with the coupling which is usually used in the analysis
of the perturbative expansions in fixed dimension $d = 3$
\cite{Nickel,LeG-ZJ}. In the large-$N$ limit another definition
 is also useful:
\begin{equation}
\widehat{g} \equiv {Ng\over 3} .
\label{hatg}
\end{equation}
Finally a fourth definition is quite common in the literature: 
\begin{equation}
f \equiv {N+2\over 3} g .
\label{deff}
\end{equation}
The quantity $f$ is naturally defined in terms of 
\begin{eqnarray}
\chi &\equiv& \int dx\, \langle \phi(0)\cdot \phi(x) \rangle, \\
\chi_4 &\equiv& \int dx dy dz\, 
    \langle \phi(0)\cdot \phi(x) \phi(y)\cdot \phi(z) \rangle_c,
\end{eqnarray}
as
\begin{equation}
f\equiv - N{\chi_4\over \chi^2\xi^d}.
\label{deff2}
\end{equation}
It is well-known that the 
$\lambda \phi^4$ model defined in Eq. (\ref{Sphi4}) 
is equivalent at criticality to the $N$-vector model defined in the 
continuum by
\begin{equation}
S= {\beta N\over 2} \int d^dx\, \partial_\mu s(x)\cdot\partial_\mu s(x)
\label{Nvectorcontinuum}
\end{equation}
where $s(x) \cdot s(x) = 1$.

On the lattice one can consider any discretization with the 
(formal) continuum limit given by Eq. (\ref{Nvectorcontinuum}). 
Here we will study the theory with nearest-neighbour interactions
defined by
\begin{equation}
S_L = -\beta N \sum_{\langle xy\rangle} s(x)\cdot s(y),
\end{equation}
where the sum extends over all lattice links $\langle xy\rangle$.
In this case the renormalized coupling constant is given 
by the previous formulae 
with the obvious substitution $\phi\to s$. Of course, on the lattice,
the integrals are replaced by sums over the lattice points.

\section{Critical-point non-analyticity in the large-$N$ limit}
\label{sec3}

An important controversial issue in the field-theory method 
at fixed dimension $d=3$ is the presence of non-analyticities 
at the critical point $g^{*}$. The question was raised
long ago by Nickel~\cite{NickelCS} who gave a simple 
argument to show that non-analytic terms should in principle be present
in the $\beta$-function. The same argument applies also to other series,
like those defining the critical exponents:
any quantity should be expected to be singular at the critical point. 

To understand the problem, let us consider the four-point
renormalized coupling $g$ as a function of the temperature $T$. For 
$T\to T_c$ we can write down an expansion of the form
\begin{eqnarray}
g = g^{*} && 
         \Bigl[ 1 + a_1 (T-T_c) + a_2 (T - T_c)^2 + \ldots + 
b_1 (T - T_c)^{\Delta} + b_2 (T - T_c)^{2 \Delta} + \ldots + \nonumber \\
&&  c_1 (T - T_c)^{\Delta+1} + \ldots + 
           d_1 (T - T_c)^{\Delta_2} + \ldots 
+ e_1 (T - T_c)^{\Delta_3} + \ldots \Bigr]
\label{grintermsTmTc}
\end{eqnarray}
where $\Delta$, $\Delta_2$, $\ldots$ are subleading exponents. 
We expect on general grounds that $a_1 = a_2 = a_3 = \ldots = 0$.
Indeed these analytic corrections arise from the non-linearity
of the scaling fields and their effect can be  
eliminated in the Green's functions by an appropriate change of variables
\cite{Fisher-Aharony-analytic}. For dimensionless
renormalization-group invariant
quantities such as $g$, 
the leading term is universal and therefore independent of 
the scaling fields, so that
no analytic term can be generated. We will explicitly show 
their absence in the following computation. 
Notice that analytic correction factors to the singular
correction terms are generally present, and
therefore the constants $c_i$ in 
Eq.~(\ref{grintermsTmTc}) are expected to be nonzero. Moreover 
we mention that, in general, the correction terms 
can include powers of the critical exponents\footnote{
In the following large-$N$ analysis we will not consider these terms.
Since for large values of $N$, $\gamma/\nu = (1-\alpha)/\nu=2$,
their contributions mix with
the analytic background, so that 
their inclusion would not change our main conclusions.}
$1-\alpha$ and $\gamma$~\cite{Fisher-Aharony-analytic}
(associated with backgrounds).

Starting from 
Eq. (\ref{grintermsTmTc}) it is easy to compute the $\beta$-function:
\begin{equation}
\beta(g) = M {dg\over dM} = {M\over dM/dT} {dg\over dT}.
\label{betadefinition}
\end{equation}
Since the mass gap $M$ scales as 
\begin{equation}
M\sim (T-T_c)^\nu
\left[1 + \bar{a}_1 (T-T_c) + ... + \bar{b}_1(T-T_c)^\Delta +
... \right],
\end{equation}
 we obtain the 
following expansion:
\begin{eqnarray}
\beta(g) &=& \alpha_1 (g^{*} - g) + \alpha_2 (g^{*} - g)^2 + 
               \ldots + 
      \beta_1 (g^{*} - g)^{1\over \Delta} + 
        \beta_2 (g^{*} - g)^{2\over \Delta} + \ldots + \nonumber \\
     && \gamma_1 (g^{*} - g)^{1+{1\over \Delta}} + \ldots +
        \delta_1 (g^{*} - g)^{\Delta_2\over \Delta} + \ldots
        +\zeta_1 (g^{*} - g)^{\Delta_3\over \Delta} + \ldots.
\label{betafunction}
\end{eqnarray}
It is easy to verify the well-known fact that
$\alpha_1 = -\Delta/\nu \equiv -\omega$ and that, if
$a_1 = a_2 = \ldots = 0$ in Eq. (\ref{grintermsTmTc}), then 
$\beta_1 = \beta_2 = \ldots = 0$. 
Eqs. (\ref{grintermsTmTc}) and  (\ref{betafunction}) are the expressions
which are expected on the basis of the standard
renormalization-group picture. They express the asymptotic behaviour
for generic models and indeed this is the supposed behaviour 
in lattice theories. However for the continuum
$\lambda \phi^4$ a much simpler expansion is
often conjectured\footnote{For a critique of these
conjectures from the point of view of the renormalization group 
\`a la Wilson, see Ref.~\cite{Sokal94}, Sec.~5.2 and
App.~E of 
Ref.~\cite{Li-Madras-Sokal}. See also Ref. \cite{Bagnuls97} for a more 
recent discussion in the same framework.}  \cite{Bagnuls85,Bagnuls90}.
First of all one assumes that the continuum theory couples only to 
one subleading scaling field, the operator associated to the 
exponent $\Delta$. As a consequence in the expansion
(\ref{grintermsTmTc}) no corrections with exponents $\Delta_2$,
$\Delta_3$, $\ldots$, should appear. In Eq. (\ref{betafunction})
this conjecture implies for instance that $\delta_1=\zeta_1=0$, i.e. 
no terms with exponents $\Delta_i/\Delta$ are present. The second claim 
is that 
$T-T_c \equiv m^2_0 - m^2_{0c}$
is a scaling field in the Wilson renormalization-group sense \cite{Wegner72}.
Therefore in Eq. (\ref{grintermsTmTc}) no term with exponent
$h\Delta + k$, $k>0$ should appear. Correspondingly in 
Eq. (\ref{betafunction}) terms of the form $h + k/\Delta$ would be 
absent. As a consequence $\beta(g)$ would be analytic. Of course,
these two hypotheses would also prove that also other expansions in $g$,
like the series for the critical exponents, would be analytic at the 
critical point.

In order to understand the validity of these conjectures, one must
compute the $\beta$-function non-perturbatively. The only case in
which this is possible is the large-$N$ limit, i.e.
in the framework of the $1/N$ expansion.

In the following we will analyze the $\beta$-function computed in 
Ref.~\cite{ONgr} for $N\to\infty$ and we will show explicitly that the 
conjecture mentioned above is incorrect: non-analytic
terms are indeed present. However, at the order of $1/N$ we are working,
we will be unable to distinguish which hypothesis is false. 
Indeed since, at $N=\infty$, 
$1 + 1/\Delta = \Delta_{2}/\Delta=\Delta_{3}/\Delta$, 
we will only be able to verify that at least one of the 
corresponding terms is present
but not to prove the presence of all of them.
This problem can only be solved by computing the next two orders in $1/N$.

The starting point is 
the $\beta$-function expanded in powers of $1/N$:
\begin{equation}
\beta(\widehat{g}) = 
   M {d \widehat{g}\over dM} = \beta^{(0)} (\widehat{g}) + 
     {1\over N} \beta^{(1)} (\widehat{g}) + O\left({1\over N^2}\right),
\end{equation}
where, for convenience, we have introduced the coupling
$\widehat{g}$ defined by Eq.~(\ref{hatg}). 
The two functions $\beta^{(0)} (\widehat{g})$ and 
$\beta^{(1)} (\widehat{g})$ were computed in Ref.~\cite{ONgr}.
One has
\begin{equation}
\beta^{(0)} (\widehat{g}) = (d-4) \widehat{g} 
   \left( 1 - {\widehat{g} \over \widehat{g}_{\infty}^{*} } \right),
\label{beta0}
\end{equation}
and 
\begin{eqnarray}
&&{\beta^{(1)}(\widehat{g})\over \widehat{g}^2}
    = (d-3)2^{d-1}\beta_0 +
    {2\over d}(d-1)^2\left( d-4+\beta_0\widehat{g}\right)^2
    \int {d^du\over (2\pi)^d} {1\over \left[1+\widehat{g}\Pi(u)\right]^2}
    {1\over \left( 4+u^2\right)^2}
\nonumber \\
&&+ 2\int {d^du\over (2\pi)^d}
    \left[ 
       {\beta_0\widehat{g} + d-4\over\left(1+\widehat{g}\Pi(u)\right)^2}
      -{\beta_0\widehat{g}\over 1+\widehat{g}\Pi(u)}
    \right]
    \left[ {1\over (1+u^2)^3}+{3\over (1+u^2)(4+u^2)}
           \left({{d\over 4}-2\over 1+u^2}-{d-1\over 4+u^2}\right)
    \right]
\nonumber \\
&&- {2\over d}\int {d^du\over (2\pi)^d}
    {(\beta_0\widehat{g} + d-4)^2\over\left( 1+\widehat{g}\Pi(u)\right)^2}
    \left[ {1\over (1+u^2)^3}+{3\over (1+u^2)(4+u^2)}\left(
    {{d\over 4}-1\over 1+u^2}+{d-1\over 4+u^2}\right)\right]
\nonumber \\
&&-4\int {d^du\over (2\pi)^d}
   \left[ 
      {\beta_0\widehat{g} + d-4\over\left(1+\widehat{g}\Pi(u)\right)^3} -
      {\beta_0\widehat{g}+{d\over 2}-2\over 
           \left( 1+\widehat{g}\Pi(u)\right)^2}
   \right]        {1\over (4+u^2)^2}
  \left( \beta_0\widehat{g} + d - 4- {3\over 1+u^2}\right)^2
\nonumber \\
&&-4\int {d^du\over (2\pi)^d}
     {\beta_0\widehat{g}(\beta_0\widehat{g} + d-4)\over
                        \left(1+\widehat{g}\Pi(u)\right)^2}
     {1\over (4+u^2)^2}
    \left( \beta_0\widehat{g} + d - 4- {3\over 1+u^2}\right).
\label{betafunctionsubunosuN}
\end{eqnarray}
Here $\widehat{g}_{\infty}^{*}$ is the critical value of $\widehat{g}$ for 
$N=\infty$
\begin{equation}
 \widehat{g}_{\infty}^{*} = {2 (4\pi)^{d/2} \over \Gamma(2-d/2)} ,
\label{defhatgstar}
\end{equation}
$\beta_0$ is defined by
\begin{equation}
\beta_0 \equiv {4-d\over \widehat{g}_{\infty}^{*}} ,
\end{equation}
and 
\begin{equation}
\Pi(u) = {1\over2} \int {d^dp\over (2\pi)^d} 
    {1\over p^2 + 1} 
   \left[ {1\over (p+u)^2 + 1} - {1\over p^2 + 1}\right].
\label{defPi}
\end{equation} 
The leading term, Eq. (\ref{beta0}), is clearly analytic. 
In App.~\ref{appunosuN} we study the 
behaviour of $\beta^{(1)}(\widehat{g})$ for 
$\widehat{g} \to \widehat{g}_{\infty}^{*}$. Setting 
\begin{equation}
\Theta \equiv {\widehat{g}_{\infty}^{*} - \widehat{g}\over
\widehat{g}}, 
\end{equation}
for $2 < d < 4$, we find
\begin{equation}
{\beta^{(1)}(\widehat{g})\over \widehat{g}^2} = 
\hbox{\rm analytic terms} + 
   A(d)  \Theta^{2\over 4-d} + \ldots,
\label{betanonanalgenerica}
\end{equation}
where 
\begin{equation}
A(d) = - \left({1\over c}\right)^{1 + {2\over 4-d} }
      {2^{d-4} (9 d^4 - 112 d^3 + 428 d^2 - 512 d + 192) \over 
        8 d (6 - d) (4-d) } 
      {\pi N_d\over \sin(2 \pi/(4-d))}, 
\label{defad}
\end{equation}
and
\begin{eqnarray}
c &=&  {\sqrt{\pi}\over2} {\Gamma(d/2 - 1) \over \Gamma((d-1)/2) }  ,
\label{defc} \\
N_d &=& {2 \over (4\pi)^{d/2}\Gamma(d/2)} .
\label{defNd}
\end{eqnarray}
This expansion is not valid whenever $d=4-2/n$, $n\in \N$, 
as in this case 
$A(d)$ diverges. For these values of the dimension one finds
\begin{equation}
{\beta^{(1)}(\widehat{g})\over \widehat{g}^2} = 
\hbox{\rm analytic terms} +
   B(n)\, \Theta^n \log\Theta + \ldots
\label{betanonanalspeciale}
\end{equation}
where
\begin{equation}
B(n) = {2\over (4-d)^2} \lim_{\epsilon\to 0} 
    \left[\epsilon\ A(d + \epsilon)\right] .
\end{equation}
In particular for $d=3$, which corresponds to $n=2$, we have
\begin{equation}
B(2) = - {71\over 12 \pi^5} .
\end{equation}
Let us now interpret the results. Let us consider first 
the case of generic $d\neq 4-2/n$.
In the large-$N$ limit the smallest exponents 
(in the region $2<d\leq 4$) are~\cite{Ma}
\begin{eqnarray}
\Delta = {4-d\over d-2} + O(1/N),\\
\Delta_{2} = \Delta_3={2\over d-2} + O(1/N).
\end{eqnarray}
At $N=\infty$ $\Delta_2$ and $\Delta_3$ are degenerate
independently of $d$~\cite{Ma}.
This degeneracy should be lifted only at the next 
order\footnote{
The corresponding scaling operators 
are linear combinations of 
$O_1\equiv (m_0^2+\case{1}{6}g_0\phi^2)(\nabla\phi)^2$ and 
$O_2\equiv (\nabla^2 \phi)^2$~\cite{Ma}.},
i.e. $O(1/N)$.
To leading order in $1/N$ one has
\begin{equation}
{\Delta_{2}\over \Delta}={\Delta_{3}\over \Delta} =
1 + {1\over \Delta} = {2\over 4 - d}.
\label{degeneracy}
\end{equation}
Eq.~(\ref{betanonanalgenerica}) shows that 
non-analytic terms are present.  But because of the degeneracy
(\ref{degeneracy}), they cannot be distinguished.
Correspondingly in Eq.~(\ref{betafunction}) one has 
\begin{equation}
\gamma_1 + \delta_1 + \zeta_1\not=0.
\end{equation}
The evaluation to $O(1/N)$ of the constant $\gamma_1$, $\delta_1$
and $\zeta_1$ requires an $O(1/N^3)$ computation.
So we cannot check if some of them are zero.
Notice that, as expected, no term
with exponent $1/\Delta$ appears in Eq.~(\ref{betanonanalgenerica})
in agreement with the argument we presented at the beginning.

Let us now consider $d=3$ (an analogous argument applies to any special
dimension $d = 4 - 2/n$). In this case the interpretation is more difficult
and we have a phenomenon analogous to the Abe-Hikami anomaly for the 
specific heat~\cite{Abe-Hikami}. The origin is an additional degeneracy
of the exponents for $N\to\infty$: the non-analytic terms 
with exponents $\Delta_2/\Delta$, $\Delta_3/\Delta$
 and $1 + 1/\Delta$ become degenerate
with the analytic term with exponent 2. This degeneracy causes the 
appearance of the logarithmic term in Eq.~(\ref{betanonanalspeciale}) 
and has the consequence that the coefficients of the non-analytic 
terms are of $O(1)$ instead of $O(1/N)$ as it was 
the case for generic values of $d$. 
Let us write each symbol entering Eq.~(\ref{betafunction}) as
\begin{equation}
\# = \#_\infty + {\#_1\over N} + O\left( {1\over N^2}\right),
\end{equation}
and expand the $\beta$-function in powers of $1/N$ to $O(1/N)$.
Then comparing with Eqs.~(\ref{beta0}) and 
(\ref{betanonanalspeciale}), one finds the relations
\begin{eqnarray}
&&\alpha_{2,\infty} + \gamma_{1,\infty} + \delta_{1,\infty} +
\zeta_{1,\infty} =
      {1\over \widehat{g}^{*}_\infty}, 
\nonumber \\
&&\delta_{1,\infty} (\Delta_{2,1} - 2 \Delta_1) +
\zeta_{1,\infty} (\Delta_{3,1} - 2 \Delta_1) - 
                   \gamma_{1,\infty} \Delta_1  =   B(2).
\label{equazionibeta}
\end{eqnarray}
The leading exponent $\Delta$ is 
known to $O(1/N^2)$ for $d=3$~\cite{Ma,omegaln}
\begin{equation}
\Delta = 1 - {32\over \pi^2}{1\over N} -
{32(9\pi^2-80)\over 3\pi^4}{1\over N^2} + O\left({1\over N^3}\right).
\label{deltaln}
\end{equation}
Eqs.~(\ref{equazionibeta}) show that at least one 
among the coefficients $\gamma_1$, $\delta_1$
and $\zeta_1$
must be non-zero, and therefore that $\alpha_{2,\infty}$,
$\gamma_{1,\infty}$,
$\delta_{1,\infty}$, and $\zeta_{1,\infty}$
are discontinuous at $d=3$. This could be a feature of the
large-$N$ limit: for finite values of $N$ 
it is still possible that all 
coefficients be continuous in $d$~\cite{Abe-Hikami-2}.

In conclusion our explicit calculation shows that non-analytic terms 
are present in the $\beta$-function defined in 
Eq. (\ref{betadefinition}). Notice that this result is valid 
for all dimensions with $2< d < 4$ and thus also in the 
$\epsilon$-expansion when one uses a massive renormalization 
scheme. In this case the singularity is of the form 
$(\widehat{g}^* - \widehat{g})^{2/\epsilon}$, a behaviour which has also been
predicted from a large-order analysis of perturbation 
theory~\cite{Parisi78e79,Bergere-David}.
However, for the $\epsilon$-expansion, it is still possible that,
as conjectured in Ref.~\cite{Schafer94}, the $\beta$-function is 
analytic in a massless renormalization scheme. The question 
requires further investigation.

For small values of $N$, i.e. $N=0,1,2,3$,  
the renormalization group analysis of Ref.~\cite{Riedel} shows that 
$\Delta_2/\Delta\simeq 2$, $\Delta_3/\Delta\simeq 3$,
and $1+1/\Delta\simeq 3$. The closeness of such values to integer
numbers may explain the small effects of the confluent non-analytic
corrections in the procedures used to estimate $g^*$.

\section{$\epsilon$-expansion results}
\label{sec4}

\subsection{Computation of ${g}^{*}$ to order 
            $O(\epsilon^4)$}  \label{sec4.1}

In this Section we will give the explicit expression of ${g}^{*}$ 
up to four loops, i.e. $O(\epsilon^4)$. There are essentially two
different methods to perform the calculation. 
One may follow the previous section: starting from the 
definition, one computes 
the renormalized two-point and four-point functions,
then derives the $\beta$-function and finally obtains 
${g}^{*}$ from the equation 
$\beta({g}^{*}) = 0$. Alternatively,
one may compute 
$g$ at three loops in terms of $g_{\overline{\rm MS}}$ and
$\epsilon$. Then, 
to express $g^{*}$ in terms of $\epsilon$ only, 
one can use the four-loop expression (i.e. $O(\epsilon^4)$)
for the fixed point value of
$g_{\overline{\rm MS}}$~\cite{fourloops}. 
We have followed this strategy,
which allows us to obtain ${g}^{*}$ to order $O(\epsilon^4)$ 
calculating only three-loop graphs. 
Some intermediate results are presented in 
App.~\ref{Appepsilon}. 
In the framework of the $\epsilon$-expansion, we found convenient
to consider the rescaled coupling $\bar{g}$, defined by 
Eq.~(\ref{gbar}). Due to the $O(\epsilon^{-1})$
factor multiplying $g$ in the definition of $\bar{g}$, 
 the $O(\epsilon^4)$ of $g^*$
corresponds to the $O(\epsilon^3)$ of $\bar{g}^*$.
The $\epsilon$-expansion of $\bar{g}^*$ to $O(\epsilon^3)$
is given by
\begin{equation}
   \bar{g}^{*}(\epsilon) = \sum_{n=0} \bar{g}_n \epsilon^n
\label{gepsseries}
\end{equation}
with
\begin{eqnarray}
\hskip -0.6truecm
\bar{g}_0 &=& 1,\\
\hskip -0.6truecm
\bar{g}_1 &=& {3(3N+14)\over (N + 8)^2} ,\\
\hskip -0.6truecm
\bar{g}_2 &=& {-2864.85 - 1086.88 N - 119.599 N^2  -
         7.07847 N^3 \over (N + 8)^4} ,\\
\hskip -0.6truecm
\bar{g}_3 &=& 
        {298347 + 165854 N + 38100.1 N^2  + 4733.16 N^3  + 240.959 N^4  +
         3.31144 N^5 \over (N + 8)^6}  .
\end{eqnarray}
We have checked, see App.~\ref{Appepsilon},
that these expressions reproduce the $O(1/N)$ results of 
Ref.~\cite{ONgr}.

\subsection{Analysis of the $\epsilon$-expansion of $\bar{g}^{*}$}

The purpose of the calculation is, of course, the determination of 
$\bar{g}^{*}$ in three dimensions and, possibly, in two dimensions.
It is well-known that one does not obtain reliable estimates 
by simply setting $\epsilon=1$ (or $\epsilon=2$) in the corresponding
series (\ref{gepsseries}), since the expansion is strongly diverging.

We have analyzed the $\epsilon$-series using the methods 
proposed in Ref.~\cite{LeG-ZJ}. The resummation technique is 
based on the knowledge of the large-order behaviour of the series.
It is indeed known  that the coefficients $\bar{g}_n$ of the series 
$\bar{g}^{*}(\epsilon)$ behave as 
\begin{equation}
   \bar{g}_n = c (-a)^n \Gamma(n + b_0 + 1) \left( 1 + O(1/n) \right).
\end{equation}
The constant $a$, which characterizes the singularity of the 
Borel transform of $\bar{g}^{*}(\epsilon)$ does not depend 
on the specific observable; it is given by~\cite{Lipatov,B-L-Z} 
\begin{equation}
 a =  {3\over N + 8}.
\end{equation}
The coefficients $c$ and $b_0$ depend instead on the series one considers.

Our analysis follows Ref.~\cite{LeG-ZJ}.
Given a quantity $R$ with series $R(\epsilon)$
\begin{equation}
R(\epsilon) = \sum_{k=0} R_k \epsilon^k ,
\label{Repsilon}
\end{equation}
we have generated new series $R_p(\alpha,b;\epsilon)$ according to
\begin{equation}
R_p(\alpha,b;\epsilon) = \sum_{k=0}^p 
    B_k(\alpha,b) 
  \int^\infty_0 dt \,e^{-t} t^b  
{u(\epsilon t)^k \over (1 - u(\epsilon t) )^\alpha}  
\label{RBorel}
\end{equation}
where 
\begin{equation}
   u(x) = { \sqrt{1 + a x} - 1\over \sqrt{1 + a x} + 1}.
\end{equation}
The coefficients $B_k(\alpha,b)$ are determined by the requirement
that the expansion in $\epsilon$ of $R_p(\alpha,b;\epsilon)$
coincides with the series (\ref{Repsilon}). For each $\alpha$, $b$ and $p$
an estimate of $R$ is simply given by $R_p(\alpha,b;\epsilon=1)$.

For $\bar{g}^{*}$
we have computed the series (\ref{RBorel}) for many values of $\alpha$ and $b$,
and for $p=2$ and $p=3$,
obtaining in this way many different estimates 
of $\bar{g}^{*}$ in three 
dimensions. We have noticed that, for $\alpha > 1$, the estimates 
strongly oscillate with the number $p$ of terms which are 
considered. These oscillations increase in size as $\alpha$ increases. 
For this reason we have 
decided to keep $\alpha$ in the interval $-1\le \alpha \le 1$.
Then for each value of $\alpha$ and $N$ we have considered various choices of 
$b$. In each case we have found an integer
 value of $b$, $b_{opt}$, such that
\begin{equation} 
R_3(\alpha,b_{opt};\epsilon=1)\approx R_2(\alpha,b_{opt};\epsilon=1).
\end{equation}
In other words $b_{opt}$ is the integer value of $b$ 
that minimizes the difference between the estimates 
from the $O(\epsilon^2)$ and $O(\epsilon^3)$ series.
In a somewhat arbitrary way 
we have then considered as our final estimate 
the average of $R_p(\alpha,b;\epsilon=1)$ with 
$-1\le \alpha \le 1$ and $-2 + b_{opt} \le b \le 2 + b_{opt}$.

Of course the real problem is the determination of the error bar. We have 
decided here to use an algorithmic procedure. The reason is that,
if error and mean value are determined algorithmically, there is less
chance  to introduce unwillingly a systematic bias due to what we expect
to be  the ``correct" value. Of course we do not only want to determine the 
error bars in a completely automatic fashion, we also want to have 
error bars which are reasonable. Since, as we will discuss below,
we will obtain many different 
estimates of $\bar{g}^{*}$ from the $\epsilon$-expansion, 
the basic requirement 
will be that all estimates should be compatible among each other. Notice 
that this is a requirement of internal consistency only, which does not use 
any external information. In this sense our procedure will be totally 
unbiased. Therefore discrepancies with results obtained from other 
methods will be meaningful.
We have thus fixed our error bar as the sum of two terms:
the first one is the variance of the values of
$R_3(\alpha,b;\epsilon=1)$
with $-1\le \alpha \le 1$ and 
$\lfloor b_{opt}/2 - 1\rfloor \le b
     \le \lceil  3 b_{opt}/2 + 1\rceil$ ; 
the second is the difference between the estimates from the 
series at order $O(\epsilon^3)$ and $O(\epsilon^2)$.

To understand the reliability of our method we have reanalyzed the 
series of the critical exponents to order 
$O(\epsilon^5)$~\cite{fourloops,fiveloops,fiveloopsb}
and compared our estimates 
and error bars with the results of Refs.~\cite{eexpLZ,eexpLZ2}. 
For $N=1$, by analyzing the series of Ref.~\cite{fiveloops}, the 
estimates $\nu = 0.6305(25)$ and $\gamma = 1.239(4)$
were obtained in Refs.~\cite{eexpLZ,eexpLZ2}.
Analyzing the same series by our method we find
$\nu= 0.6272(32)$ and $\gamma = 1.2343(35)$.
These estimates agree within errors and also 
the error bars are very similar. Our 
algorithmic procedure appears to give results similar to the 
estimates of other authors\footnote{ 
It should be noticed that the 
series of Ref.~\cite{fiveloops} contained an error in the 
five-loop coefficients as shown in Ref.~\cite{fiveloopsb}. 
An analysis of the correct series has never
been published.
In Ref.~\cite{fiveloopsb} the authors only mention that the
analysis of the correct series gives results  that are
consistent
with those of Refs.~\cite{eexpLZ,eexpLZ2}.
For the sake of completeness, we performed
a new analysis of these series using our procedure.
For $N=0,1$ we employed a constrained analysis using
the known exact results 
in two dimensions, as in Refs.~\cite{eexpLZ,eexpLZ2}.
We found
$\nu = 0.5882(11)$ and
$\gamma = 1.1559(10)$ for $N=0$;
$\nu = 0.631(3)$
and $\gamma = 1.240(5)$ for $N=1$.
We mention that perfectly consistent results are obtained
using also the homografic transformation 
$\epsilon'=\lambda\epsilon/(\lambda-\epsilon)$~\cite{LeG-ZJ}
with $\lambda=4,3$ for $N=0,1$ respectively.
We also report results from an unconstrained analysis
of the series for $N=2,3$:
$\nu = 0.664(3)$ and $\gamma = 1.304(7)$ for 
$N=2$; $\nu = 0.699(4)$ and $\gamma=1.372(6)$  for $N=3$. 
For $N=2,3$ this analysis seems to underestimate 
(slightly) the values of the exponents
(see e.g. Refs.~\cite{ZJbook,B-C}).
Indeed by employing a homografic transformation
with $\lambda=2$ (or constraining the exact two-dimensional
values: $1/\nu=1/\gamma=0$) one obtains larger values
by approximately two per cent
($\nu\simeq 0.676$ and $\gamma\simeq 1.324$ for $N=2$,
and $\nu\simeq 0.712$ and $\gamma\simeq 1.394$ for $N=3$),
but with much larger ``errors''.
\label{expfn}}.

It has been noticed that the estimates of the 
$\epsilon$-expansion can be improved if one knows the (exact) value
of the quantity one is considering in two dimensions~\cite{eexp}
or even in one dimension~\cite{DesClo}.
We will now try to do more. First of all we expect 
$\bar{g}^{*}(\epsilon)$ to be analytic 
in the domain $0< \epsilon < 4$. This conjecture 
can be checked in the large-$N$ limit using the exact results of 
Ref.~\cite{ONgr}. Moreover it has been implicitly assumed in the 
dimensional expansion around $d=0$ done
in Refs.~\cite{Bender-etal,Bender-Boettcher}.
Then we will try to improve our estimates 
using the exactly known results for $\epsilon=3$ 
($d=1$) and $\epsilon=4$ ($d=0$)
and furthermore the estimates for $\epsilon = 2$ ($d=2$). 

Let us now discuss the procedure which is a simple generalization 
of the technique presented in Ref.~\cite{eexp}. 
Suppose that 
for $\epsilon = \epsilon_1$ the exact value $R_{\rm ex}(\epsilon_1)$ is 
known. One may then define 
\begin{equation}
\overline{R}(\epsilon) = \left[ 
    {R(\epsilon) - R_{\rm ex}(\epsilon_1) \over (\epsilon - \epsilon_1)}\right]
\end{equation}
and a new quantity 
\begin{equation}
   R_{\rm imp}(\epsilon) = R_{\rm ex}(\epsilon_1) + 
         (\epsilon - \epsilon_1) \overline{R}(\epsilon) .
\end{equation}
New estimates of $R$ at $\epsilon = 1$ can then be obtained by 
applying the resummation procedure we described above to 
$\overline{R}(\epsilon)$  and then computing 
$R_{\rm imp}(1)$. 
This strategy can be generalized to an 
arbitrary number of points: if exact values 
$R_{\rm ex}(\epsilon_1)$, $\ldots$, $R_{\rm ex}(\epsilon_k)$ are known 
for a set of dimensions $\epsilon_1$, $\ldots$, $\epsilon_k$, 
$k \ge 2$, then one defines 
\begin{equation}
Q(\epsilon) = \sum_{i=1}^k \left[
    {R_{\rm ex}(\epsilon_i) \over (\epsilon - \epsilon_i)}
    \prod_{j=1,j\not=i}^k (\epsilon_i - \epsilon_j)^{-1} \right]
\end{equation}
and 
\begin{equation}
\overline{R}(\epsilon) = 
   {R(\epsilon) \over \prod_{i=1}^k (\epsilon - \epsilon_i)} - 
   Q(\epsilon) ,
\end{equation}
and finally
\begin{equation}
   R_{\rm imp}(\epsilon) = \left[ Q(\epsilon) + 
      \overline{R}(\epsilon) \right]
      \prod_{i=1}^k (\epsilon - \epsilon_i)  .
\label{serieconstrained}
\end{equation}
One can easily verify that the expression
\begin{equation}
   \left[ Q(\epsilon) + 
      \overline{R}(0) \right]
      \prod_{i=1}^k (\epsilon - \epsilon_i)  
\label{interpolation}
\end{equation}
represents the $k$-order polynomial interpolation 
among the points $\epsilon=0,\epsilon_1,...,\epsilon_k$.
Again the resummation procedure is applied to $\overline{R}(\epsilon)$
and the final estimate is obtained computing $R_{\rm imp}(1)$.
The idea behind this method is very simple. If, for instance, the value 
of $R$ for $\epsilon = 2$ is known, one uses as a zeroth order 
approximation at $\epsilon = 1$ the value of the linear interpolation 
between $\epsilon = 0$ and $\epsilon = 2$ and then uses the series 
in $\epsilon$ to compute the deviations. If the interpolation is a 
good approximation one should find that the series which gives the 
deviations has smaller coefficients than the original one. 
Consequently also the errors in the resummation are reduced.
In our case the value of $\bar{g}^{*}$ is known for $\epsilon = 3$ and 
$\epsilon = 4$. For $\epsilon=2$ we will use estimates obtained 
using strong-coupling methods or, for larger values of $N$, 
derived from the $1/N$ expansion.
Using this additional information we will be able to 
substantially reduce the error on our estimates.

As an example of the procedure let us consider the Ising model, $N=1$.

In this case the series of $\bar{g}^{*}(\epsilon)$ is given by
\begin{equation}
\bar{g}^{*}(\epsilon) = 
   1 + 0.629629 \epsilon - 0.621613 \epsilon^2 + 0.954535 \epsilon^3 
+ O(\epsilon^4) .
\label{seriesN=1}
\end{equation}
The coefficients  are big, and at first sight it may seem hopeless 
to try to get an estimate of $\bar{g}^{*}$ for $\epsilon = 1$.
However the series alternates 
in sign and therefore one may obtain reasonable 
results after a Borel
transformation. Indeed we find $\bar{g}^{*} = 1.37 \pm 0.09$
which is already quite good. We can now try to improve our 
estimates using the value of $\bar{g}^{*}$ for 
$\epsilon = 2$. Using the two-dimensional estimate
$\bar{g}^{*} = 1.7540(2)$ (see next Section)
and Eq. (\ref{serieconstrained}),
we get a new series for $\bar{g}^{*}$:
\begin{equation} 
 \bar{g}^{*}_{\rm imp}   =
      1.7540 + \Delta g (1)  ,
\end{equation}
where 
\begin{equation}
\Delta g (\epsilon) = - 0.37698 + 0.12632 \epsilon - 
           0.24765 \epsilon^2 + 0.35345 \epsilon^3 
+ O(\epsilon^4) .
\end{equation}
The coefficients of the new series are on average a factor of three 
smaller than the coefficients of the original series and 
the simple interpolation already gives a good estimate of 
$\bar{g}^{*}_{\rm imp}$: indeed $1.754 + \Delta g (0) = 1.377$,
not very far from the correct value.  Consequently 
one expects a corresponding gain in the error bar. Indeed we get 
$ 1.400 \pm 0.017 $. There is also an additional error due to the 
uncertainty in the two-dimensional result which in this case however
turns out to be negligible. One can go further 
and use both the estimate for $\epsilon = 2$ and the exact result 
for $\epsilon = 3$~\cite{ONgr}, which is $\bar{g}^{*} = 9/4$. 
Following the procedure we presented above, we get a new estimate 
of $\bar{g}^{*}$ from 
\begin{equation} 
 \bar{g}^{*}_{\rm imp}   =
      1.2580 + \Delta g (1) .
\end{equation}
where 
\begin{equation}
\Delta g (\epsilon) = 
  0.07937 + 0.110671 \epsilon - 0.128206 \epsilon^2 + 0.192895 \epsilon^3 
+ O(\epsilon^4) .
\label{deltag12}
\end{equation}
On average the coefficients of the new series are a factor of two smaller
than those of the series in which only the estimate at $\epsilon = 2$
was used. Correspondingly we obtain a slightly more precise 
estimate $\bar{g}^{*} = 1.395 \pm 0.016$. Again the error 
due to the uncertainty on the two-dimensional result is negligible.

Finally we can include also the known value of $\bar{g}^{*}$
for $\epsilon = 4$, $\bar{g}^{*} = 3$. 
Using the values at $\epsilon=2,3,4$, we get a new expansion in the 
form
\begin{equation} 
     \bar{g}^{*}_{\rm imp}   =
      1.514 + \Delta g (1) .
\end{equation}
where 
\begin{equation}
\Delta g (\epsilon) = 
  - 0.13095 +  0.05027 \epsilon - 0.08359 \epsilon^2 + 0.12377 \epsilon^3 
+ O(\epsilon^4) .
\end{equation}
Apart form the first term of the series, all the other terms have coefficients 
which are 
smaller than those of the series (\ref{deltag12}) and thus we expect
an additional reduction of the error. Indeed we find 
$\bar{g}^{*} = 1.397 \pm 0.008$. 

One can try other possibilities, constraining the 
series only at $\epsilon = 3$ or $\epsilon = 4$, or at any 
possible pair. In  Table~\ref{epsilon3d} we present also the results 
obtained by constraining the series at $\epsilon = 3$ and 
at $\epsilon = 3,4$.
The estimates are all consistent among each other, thus giving 
confidence to the final result.

We have applied this method to all values of $N$. In one dimension
$\bar{g}^{*}$ was computed in Ref.~\cite{ONgr} finding
\begin{equation}
\bar{g}^{*}=
    {N+8\over N+2} \left(1 - {1\over 4 N}\right)
\qquad\qquad {\rm for} \quad N\ge 1,
\label{gstar1da}
\end{equation}
and
\begin{equation}
\bar{g}^{*} =     {N+8\over 4} 
\qquad\qquad {\rm for} \quad N\le 1.
\label{gstar1db}
\end{equation}
In zero dimensions we generalize the result of Ref.~\cite{Bender-Boettcher} 
to any integer $N\ge 1$. We get
\begin{equation}
\bar{g}^{*} = {N+8\over N+2}  \qquad \qquad
{\rm for} \quad N\ge 1.
\label{gstar0d}
\end{equation}
It is not clear how to obtain the value of $\bar{g}^{*}$ for $N=0$. 
The one-dimensional results, Eqs.~(\ref{gstar1da}) and 
(\ref{gstar1db}), indicate that
one cannot naively set $N=0$ in the formula obtained for $N\ge 1$. 
The incorrectness of this analytic continuation can also be 
verified by our constrained analysis: if one uses the prediction 
$\bar{g}^{*} = 4$
for $\epsilon = 4$ one obtains results which are in total disagreement
with the other estimates.

In $d=2$ we will use the best available estimates. 
For $N\le 4$ we consider the estimates obtained from the 
analysis of the strong-coupling series, more precisely,
$\bar{g}^{*} = 1.679(3)$, 1.7540(2), 1.810(10), 1.724(9), 1.655(16)
respectively for $N=0,1,2,3,4$ (see next Section). For 
$N\ge 8$ we use the large-$N$ expression~\cite{ONgr}
\begin{equation}
\bar{g}^{*} = {N + 8\over N + 2} 
   \left(1 - {0.602033\over N}\right) .
\label{gstarlargeN2d}
\end{equation}
Of course the prefactor in front of the previous expression is 
arbitrary. However if one uses this particular form 
one finds a small $1/N$ correction and good agreement up to $N=4$.
Indeed comparison with the (imprecise) strong-coupling result 
at $N=4$ we have reported above shows that 
Eq. (\ref{gstarlargeN2d}) reproduces 
the correct result with an error smaller than 4\%. 
In the absence of better estimates for $N > 4$ we have thus used 
formula (\ref{gstarlargeN2d}) with an ``estimated"
error of $0.64 \bar{g}^{*}/N^2$ (which reproduces
the difference of 4\% found at $N=4$).

The final results of our analysis for selected values of $N$ 
are reported in Table~\ref{summary3d}.

Our results can be checked in the large-$N$ limit using the exact result,
$\overline{g}^* = 1 + 4.4540/N$ \cite{ONgr}. Let us consider the 
constrained analysis in $d=0,1,2$, that is the one that provides
the most precise estimates. In this case $\overline{g}^*$ 
is determined from 
\begin{equation}
\overline{g}^* = 1 + {4.9439\over N} + 
   \Delta g(1),
\end{equation}
where
\begin{equation}
\Delta g(\epsilon) = 
   {1\over N} (- 1.09695 + 0.90986 \epsilon - 
     0.359826 \epsilon^2 + 0.051144 \epsilon^3).
\end{equation}
Applying our resummation procedure to $\Delta g(\epsilon)$ we obtain
the estimate 
\begin{equation}
\overline{g}^{*} = 1 + {4.448 (10) \over N}
\end{equation}
in perfect agreement with the exact result reported above. 

We have also repeated the analysis in two dimensions. In this case of course
it is more difficult to get precise estimates: the unconstrained 
expansion gives results with errors of order $20-50$\% and it is therefore
practically
useless. Better estimates are obtained constraining the expansion
in one and zero dimensions. The results for these two cases are reported in 
Table~\ref{epsilon2d}, and are consistent.
Of course the errors are larger than in three dimensions,
but still small if one considers the series we started from 
(for $N=1$ see Eq. (\ref{seriesN=1})). The resulting estimates
of $\bar{g}^*$  are in good 
agreement with the 
strong-coupling and large-$N$ estimates we used above 
in the analysis of the three-dimensional case, thus supporting their
use.

\section{Strong-coupling expansion}
\label{sec5}

In order to understand the reliability of our previous estimates for the 
renormalized coupling constant, it is useful to compare them
with the results of a quite different approach, such as
the analysis of the high-temperature expansion for
the $N$-vector model.
In this context it is convenient to consider
the coupling $f$ defined in Eqs.~(\ref{deff}) and (\ref{deff2}).
The finiteness of $f^*$ is related to the hyperscaling
relations of the critical exponents.
A detailed analysis of the strong-coupling expansion for
all values of $N$ was presented in Ref.~\cite{ONgr} 
(see also Ref.~\cite{Reisz})
extending the  numerous studies for the 
Ising model~\cite{Bakerold,Nickel-Sharpe,Bakergr2,B-etal,Z-L-F,B-C-g}.
The results were in substantial agreement with the renormalization-group
estimates, except for small values of $N$, where 
relatively small systematic deviations 
were found. The reason of these discrepancies is the presence 
of confluent singularities at $\beta_c$. 
Indeed, in general $f(\beta)$ behaves as
\begin{equation}
f(\beta) = f^* + c_\Delta (\beta_c - \beta)^\Delta + \ldots 
\label{conf}
\end{equation}
close to the critical point.
The traditional methods of
analysis, like Pad\`e and Dlog-Pad\`e approximants,
are unable to handle an asymptotic behaviour 
like (\ref{conf}) when $\Delta$ is not an integer number, 
thus leading to a systematic error.

To take into account this kind of confluent corrections in full generality, 
one should consider 
integral approximants, which, however, require long series to 
detect non-leading effects,
and in practice need to be biased to work well.
Roskies proposed a rather simple method to handle the leading
confluent singularity in the Ising model~\cite{Roskies},
where $\Delta\simeq 1/2$. He showed that
the effect of the non-analytic terms
can be significantly reduced  by a suitable change of variables.
Equivalently one can use suitably biased integral approximants
\cite{B-C}. In Ref.~\cite{B-C-g} the series for the renormalized 
coupling constant for the Ising model was analyzed 
biasing $\Delta=1/2$.
The estimated value of $f^{*}$ was significantly lower 
than previous estimates, and now in good agreement with the 
renormalization-group prediction.
This procedure was  also successfully applied to the 
calculation of the critical exponent of the specific heat
from the low-temperature  expansion~\cite{Guttmann-R-T},
providing results consistent with field theory, while a
standard analysis neglecting confluent singularities 
led to  a quite inconsistent estimate~\cite{Bhanot}.

We have decided to repeat the analysis of Ref.~\cite{ONgr} 
for all values of $N$ using 
similar ideas. The new results show systematic differences from the old 
ones and now they are in much better agreement
with the renormalization-group estimates.

The strong-coupling expansion of $f(\beta)$ has the following form
\begin{equation}
f(\beta)= 
{1\over \beta^{d/2}} \sum_{i=0}^\infty a_i \beta^i .
\label{cha}
\end{equation}
In three dimensions 
the available strong-coupling series allow us to calculate
\begin{equation}
A(\beta)\equiv\beta^{3/2}f(\beta)
\label{adef}
\end{equation}
up to $14^{\rm th}$ order
(using $\chi$ and $m_2$ up to 15th order~\cite{B-C,d3paper},
and $\chi_4$ up to 14th order~\cite{Luscher,ButeraON}).
Longer series are available for the Ising model, $N=1$.
On the cubic lattice, using the 
published series~\cite{K-Y-T,McKenzie,B-C-g}, one can
derive $A(\beta)$ up to 16th order.
Moreover series on other lattices are 
available,  which allow us to calculate $A(\beta)$ 
on the b.c.c., f.c.c. and diamond
lattice, up to 13th, 10th,
and 19th order respectively (using series
published in Refs.~\cite{K-Y-T,McKenzie,NickelCS,M-J-W,d3paper}).
We mention that longer series for all values of $N$ 
on the cubic and b.c.c. lattice
have been announced in Refs.~\cite{Reisz,B-C},
but they have not been published yet. 

The idea of the Roskies transform (RT)~\cite{Roskies} is to 
perform biased analyses which take into account the leading confluent
singularity. For the Ising model,
where\footnote{We mention 
the recent estimate of $\Delta$ for $N=1$ obtained by the fixed-dimension
field-theoretic method: $\Delta=0.498(8)$~\cite{G-Z}.
Notice that this is somewhat lower than
the high-temperature estimate~\cite{C-F-N,Z-F}
$\Delta=0.54(3)$.\label{fdn1}}  $\Delta\simeq 1/2$, 
one replaces the variable $\beta$ 
in the original expansion with a new variable $z$, defined by
\begin{equation}
1-z = (1 - \beta/\beta_c)^{1/2}.
\label{RT}
\end{equation}
Of course a quite precise estimate of $\beta_c$ is required
here.
If the original series has square-root correction terms,
the transformed series has analytic correction terms,
which can be handled by standard 
Pad\`e or Dlog-Pad\`e approximants.
 
In order to analyze models with different values
of $N$ and therefore with $\Delta\neq 1/2$,
one can generalize the Roskies transformation and consider
the change of variable~\cite{A-M-P}
\begin{equation}
1-z = (1-\beta/\beta_c)^b.
\label{cRT}
\end{equation}
In the following we will refer to this trasformation as GRT.
For $b=\Delta$, this mapping makes the first correction
to scaling analytic. Thus standard approximants can handle it correctly.
Of course non-analytic terms still survive due to subleading corrections,
but they should be less important as far as
$\Delta$ is sufficiently smaller than the next exponents. 
Anyway they still represent a (hopefully small) source
of systematic error for our analysis. 
From Eq.~(\ref{cRT}) it follows that
\begin{equation}
f(\beta)\longrightarrow \bar{f}(z) \equiv z^{-3/2}\bar{A}(z).
\label{barA}
\end{equation}
In order to estimate $f(\beta_c)=\bar{f}(1)$
we analyze the expansion of $\bar{A}(z)$  in powers of $z$.
As we shall see, the use of the mapping (\ref{cRT})
leads to a much better agreement with 
field-theoretic estimates.

Note that the relevant singularity of $\bar{f}(z)$, i.e. the
one at $z=1$, is no longer the closest to the origin.
Indeed, the antiferromagnetic singularity at $-\beta_c$ is mapped
closer to the origin than that at $\beta_c$, but still at negative values
of $z$. We expect its effect to be small when
evaluating the resummed series around $z=1$.

By employing standard resummation methods, we studied
the behavior of $\bar{f}(z)$ around $z=1$. 
For the sake of comparison,
we also performed a standard analysis, i.e. without
using the mapping (\ref{cRT}).
We constructed various types of approximants to the series
of $\bar{A}(z)$,
such as Pad\'e approximants (PA's), Dlog-Pad\'e approximants (DPA's) 
and first-order inhomogeneous 
integral approximants (IA's)
(for a review on the resummation techniques
see for example Ref.~\cite{Guttrev}).
Note that in principle IA's should be  able to detect 
the first non-analytic correction to scaling in Eq.~(\ref{conf}),
but they probably need more terms of the series, and
practically need to be explicitly biased as in the case of PA's and
DPA's. Indeed the IA results without the GRT turn out to be
 substantially equivalent to those obtained from PA's and DPA's. 
In all cases we considered only quasi-diagonal
approximants\footnote{
Given a $n$th order series,
we considered the following quasi-diagonal approximants:
$[l/m]$ PA's and DPA's with
$l+m \geq n-2$ and
$l,m \geq  {n\over 2}-2$
($l,m$ are the orders of 
the polynomials respectively in the numerator and denominator of the PA
of the series in the case of PA's, or 
of its logarithmic  derivative in the case of DPA's);
$[m/l/k]$ IA's  with $m+l+k+2=n$ and
$\lfloor (n-2)/3 \rfloor -1\leq m,l,k \leq \lceil (n-2)/3\rceil +1$
($m,l,k$ are the orders of the polynomial $Q_m$, $P_l$
and $R_k$ defined by the first-order linear differential equation
$Q_m(x)f^\prime (x)+ P_l(x)f(x)+R_k(x)= O\left( x^{k+l+m+2}\right)$,
whose solution provides an approximant of the series at hand).
\label{ff}}. 
Then we evaluated the approximants  at $z=1$
in order to obtain an estimate of the fixed-point value of $f$. 
Very precise values of $\beta_c$, which are needed for the mapping
(\ref{cRT}), are available
in the literature from different calculations
(see for example
Refs.~\cite{MacDonald,Grassberger,Blote,Hasenbush,Chen-Holm,B-C,ONgr}).
Errors due to
the uncertainty on the value of $\beta_c$ turned out to be negligible
in our analysis.
As estimates of $\Delta$ we used the
field-theoretic prediction for $N\le 24$, and its 
large-$N$ expression (\ref{deltaln}) for larger values of $N$.
Since at $N=0$ there is a relatively large
difference between the field-theoretic prediction, $\Delta=0.470(25)$,
and the Monte Carlo estimate, 
$\Delta=0.515(7)^{(+10)}_{(-0)}$~\cite{Nickellast},
in our analysis
we considered the very conservative value 
$\Delta=0.50(5)$. Similarly for $N=1$ (see footnote
\ref{fdn1}) we consider the value $\Delta=0.50(5)$.

In a PA or DPA  analysis of quantities with a confluent 
non-analytic correction,
the singularity should be mimicked by 
shifted poles at $\beta\gtrsim \beta_c$. 
PA's and DPA's of  $A(\beta)$ present indeed
singularities typically at  $\beta\simeq 1.1\div 1.2\;\beta_c$.
On the other hand, 
the approximants of the series in $z$
(cf. Eq.~(\ref{cRT}))
do not show singularities close to the new critical value $z=1$,
conferming the effectiveness of this change of variable.  
Most approximants of $\bar{A}(z)$ do not present 
singularities near the real axis in the region ${\rm Re}\ z < 1.5$.

Table~\ref{SCresults} shows the results of our strong-coupling
analysis\footnote{ As estimate of $f^*$ 
from each class of approximants (i.e. PA's, DPA's, and IA's) 
we took the average of the 
values at $z=1$ (or $\beta=\beta_c$ when performing a standard analysis)
of the non-defective approximants 
using all the available terms of the series. 
The error we quote is 
the square root of the variance around the estimate of the results 
from all the non-defective approximants listed in the footnote \ref{ff}.
Approximants in the generic variable $x$
are considered defective when
they present spurious singularities close to the real axis for ${\rm
Re}\ x \lesssim x_c$.
The results from PA's, DPA's, and IA's are then combined
leading to the estimates shown in Table~\ref{SCresults}.}
for selected values of $N$.
There we report also the values of $\beta_c$ and $\Delta$ we used.
The results of the analyses which use the GRT are quoted with two errors:
the first one is the spread of the approximants for $b=\Delta$,
the second one is due to the uncertainty on $\Delta$, and it
is obtained by varying $b$.
We note that for $N=0,1,2$ the estimate of $f^*$  
increases with increasing $b$ in the GRT.

A few comments on the results of Tables~\ref{SCresults} are in order.

(a) For most values of $N$,
the difference between the results of the analysis with and without the 
use of the GRT, although relatively small, is  
larger than the apparent error
of the single analysis (which is estimated by looking at the stability 
of the different approximants).
This indicates that the systematic error
due to the neglecting of confluent singularities is much larger than
the error obtained by a stability analysis of the results.

(b) The estimates from the
GRT analyses are globally
in much better agreement with the field-theoretic predictions
than the results obtained neglecting the confluent singularities 
(except for $N=4$, for which the agreement 
of the non-biased result was already satisfactory).
This can be seen in Table~\ref{summary3d} where
the corresponding results for $\bar{g}^*$ are reported.
We note that the strong-coupling estimates are systematically
slightly higher for $N\geq 3$. 
The  uncertainty of the GRT results is approximately 1\%,
thus providing an accurate check of the field-theoretic calculations
by a different approach.

(c) For the Ising model, universality 
among formulations on the cubic, b.c.c., f.c.c., and diamond
lattices is well
verified by the results of our GRT analysis.
Assuming universality, 
our overall estimate is $f^*=23.55(15)$. 
This is consistent with the result of 
the biased analysis (sligthly different from ours)
of the strong-coupling expansion on the cubic lattice
of Ref.~\cite{B-C-g}, $f^*=23.69(10)$.
We note that universality is apparently shown
also by the results of the standard analysis, although they lead to 
a different estimate of $f^*$.
This may be explained by noting that
if the values of the non-universal coefficients $c_\Delta$ 
in Eq.~(\ref{conf})
are approximately the same for all the considered lattice formulations,
they may give rise to similar systematic errors leading
to an apparent universality of the results.
This would not be surprising. Indeed, 
already the leading amplitudes of many non-universal
quantities have close values in various
nearest-neighbor lattice formulations
(see e.g. the results reported in Ref.~\cite{Z-L-F}, 
and the estimates 
of some amplitudes of leading scaling corrections reported in Ref.~\cite{L-F}).  
In Ref.~\cite{Z-L-F}
a different analysis still neglecting confluent singularities (where
the amplitudes of $\chi$, $m_2$ and $\chi_4$ were indipendently
calculated by a
first order integral approximant analysis to give an estimate to $f^*$)
led to the following results:
$f^*=24.55^{+0.95}_{-0.25}$ on the cubic lattice,
$f^*=24.39(9)$ on the b.c.c. lattice, and
$f^*=24.50(13)$ on the f.c.c. lattice.
Universality is nicely observed, but, again, the final result is 
larger than the GRT estimate, and therefore also
than the field-theoretic ones.
Notice that these estimates of $f^*$ are lower than
our results obtained without using
the GRT. This is probably due to the 
different procedure used to
estimate $f^*$, and 
to the fact that different series were analyzed. 

(d) With respect to the standard analysis, the GRT results
are in much better agreement with the
formula obtained by a $1/N$ expansion~\cite{ONgr}: 
\begin{equation}
f^* =  16\pi \left[ 1 - {1.54601\over N}+O\left( {1\over N^2}\right)
\right].
\label{fstarln}
\end{equation}
This equation gives:
$f^*=48.646$ for $N=48$,
$f^*=47.837$ for $N=32$,
$f^*=47.027$ for $N=24$,
$f^*=45.408$ for $N=16$, etc...
This agreement provides additional support to the formal argument presented
in Sec.~\ref{sec3}, according to which analytic terms should not be present
in the expansion of $f$ around $\beta_c$. Indeed, in the opposite case,
one would not expect a substantial improvement 
using the mapping (\ref{cRT}) for $N$ sufficiently large,
since $\Delta\rightarrow 1$ for $N\to\infty$.

We mention that for the Ising model
high-temperature techniques have also been
used to obtain a dimensional expansion 
of the Green's functions around  $d=0$.
The analysis of these series presented in Ref.~\cite{Bender-Boettcher} led 
to the quite good estimate $f^*=23.66(24)$.

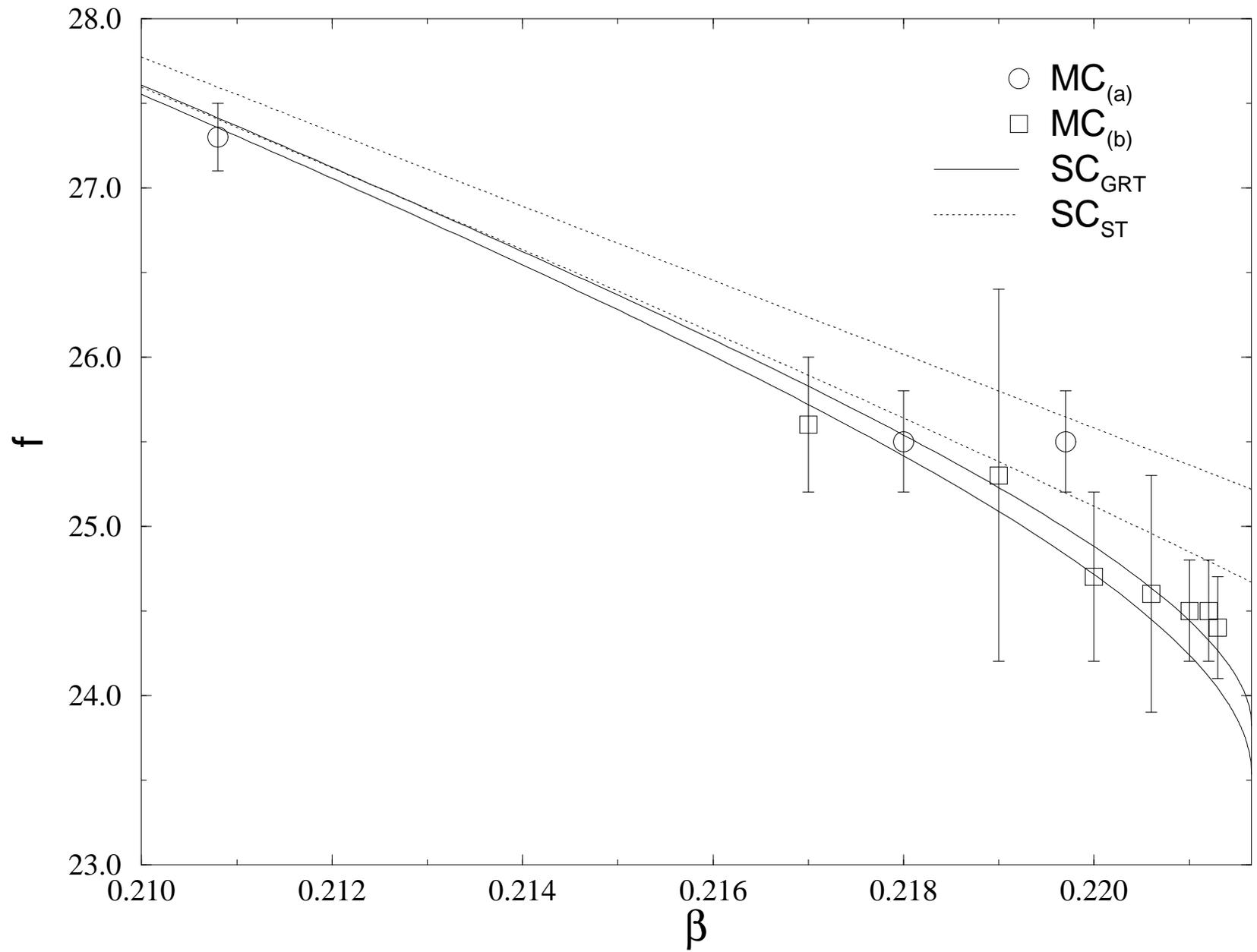
\begin{figure}
\caption{
Plot of the function
$f(\beta)$ of the 
three-dimensional Ising model as
obtained  from the quasi-diagonal PA's calculated
with and without the use of the GRT
(for each case we draw two lines representing the corresponding
band of uncertainty). 
For comparison some Monte Carlo data taken
from Refs.~\protect\cite{BakerKawa} ($a$) and \protect\cite{K-L} ($b$)
are also plotted. 
}
\label{fig1}
\end{figure}

Let us compare our strong-coupling predictions
with the results obtained from Monte Carlo simulations.
Monte Carlo estimates of $f^*$ for the Ising model on the cubic lattice
can be found in 
Refs.~\cite{Freedmanetal,Freedman-Baker,Wheater,Weston,K-P,Tsypin,BakerKawa,K-L}.
In Fig.~\ref{fig1} 
we compare some of the data of the most recent works with the
Pad\`e resummations of the strong-coupling series with and without the GRT.
The data of Ref.~\cite{K-L},
which are those closest to criticality, have been obtained
by employing a finite-size-scaling technique,
which allowed the authors to get data up to a value of $\beta$ corresponding 
to $\xi\simeq 30$.
These data show $f(\beta)$ apparently flattened around $24.5(2)$.
Ref.~\cite{BakerKawa} presents data up to $\xi\simeq 10$, 
from which the authors obtain the value $f^*=25.0(5)$. 
Fig.~\ref{fig1} shows that
Monte Carlo data are in substantial
agreement with our GRT analysis\footnote{
One should also take into account that the
data of Ref.~\cite{K-L} at different $\beta$'s
are not statistically independent
because they have been obtained by a finite-size scaling technique.}.
But biased approximants extrapolate to a smaller value of $f^*$.
It is worth mentioning that the use of the GRT to bias
the strong-coupling approximants is, in a sense,
equivalent to the use of the function
\begin{equation}
f(\beta) = f^* + c_\Delta(\beta_c-\beta)^{1/2}
\label{fit}
\end{equation}
for the extrapolation to $\beta_c$ of the Monte Carlo data at
$\beta<\beta_c$. Using the function (\ref{fit}) to fit
the Monte Carlo data of Fig.~\ref{fig1} one gets
$f^* = 23.7(2)$ and $c_\Delta=34(3)$
with $\chi^2/{\rm d.o.f}\simeq 0.6$
(here we assumed all data to be independent)\footnote{
If we do not include the data for the lowest value of $\beta$,
corresponding to $\xi\simeq 3$, we obtain $f^*=23.9(3)$ 
and $c_\Delta=27(7)$.},
which is perfectly
consistent with the strong-coupling and field-theoretic estimate
of $f^*$. Finally we mention the result of Ref.~\cite{Tsypin}:
$f^*=23.3(5)$, obtained by studying the probability
distribution of the average magnetization.

We mention another application of the
Roskies transform (\ref{RT}), that is 
the analysis of the low-temperature expansion of the quantity
$u$ defined in the broken phase of the Ising model by 
\begin{equation}
u\equiv {3\chi\over \xi^3 M^2 },
\end{equation}
where $M$ is the magnetization.
$u$ plays the important role of a 
zero-momentum low-temperature renormalized
coupling constant in the study of the $\phi^4$ theory
directly in $d=3$~\cite{Munster}.
The most precise determinations up to now have been apparently
obtained by Monte Carlo simulations~\cite{H-C}
(where data have been fitted by using Eq.~(\ref{fit})):
$u^*=14.3(1)$,
and by two different
analyses of the low-temperature expansion~\cite{Siepmann,Z-L-F}, 
which lead to  apparently inconsistent results:
$u^*=14.73(14)$~\cite{Siepmann} and 
$u^*=14.14(14)$~\cite{Z-L-F,Fisher-pri}.
We repeated the analysis of the low-temperature
expansion using the RT.
The series published in 
Refs.~\cite{Ar-Ta,Vohw} allow us to calculate the
expansion of $u$ in powers of $e^{-4\beta}$ up to
21th order. Quasi-diagonal PA's of the Roskies-transformed
series give
$u^*=14.3(1)$ (without using the RT one obtains $u^*=14.7(1)$ as
in Ref.~\cite{Siepmann}), thus confirming
the Monte Carlo result of Ref.~\cite{H-C} and the low-temperature
analysis of Ref.~\cite{Z-L-F}.

In Sec.~\ref{sec4} we
used estimates of $f^{*}$ in two dimensions
for our constrained analysis of the $\epsilon$-expansion.
In the following we shortly discuss their derivations.
An analysis
with Pad\`e approximants of a 17th-order series was presented
by Butera and Comi \cite{B-C-2d}. For $N\le 2$ they found:
$f^{*}=10.53(2)$ for $N=0$,
$f^*=14.693(4)$ for $N=1$, and 
$f^*=18.3(2)$ for $N=2$.
We reanalyzed the same series using also DPA's and IA's.
For $N=2$ we  also
considered the series in the internal energy
$E$~\cite{ONgr}. We found
results in total agreement: $10.55(2)$ for $N=0$, $14.693(1)$ for $N=1$
and $18.2(1)$ for $N=2$. 
Moreover, 
using the series for the Ising model on the triangular lattice
published in Refs.~\cite{K-Y-T,McKenzie,ON-d2-b},
we obtained the strong coupling expansion of $\beta f(\beta)$ 
to 14th order. Its analysis gave $f^*=14.695(1)$.
A comparison with the square-lattice result 
(i.e. assuming universality) leads to the
final estimate $f^*=14.694(2)$ for the two-dimensional Ising model.

For $N=3$ no estimate was reported in Ref.~\cite{B-C-2d}. Our analysis gives
\begin{equation}
f^* = 19.7(1) \qquad \hbox{\rm for $N = 3$.}
\label{fstarN3}
\end{equation}
For comparison we mention the field-theoretic estimate
obtained in Ref.~\cite{Falcionietal} for $N=3$: $f^*=20.0(2)$.

Estimates for several values of $N > 3$ were reported in Ref.~\cite{B-C-2d}.
For $N=4$ $f^* = 20.9(1)$ was found, which is 
in good agreement with our reanalysis: $f^* = 20.8(2)$. 
For larger values 
of $N$ --- we will be interested in $N\ge 8$ ---
the analysis of the strong-coupling series gives 
results with a somewhat large error and in this case 
the $1/N$ expression~\cite{ONgr}
\begin{equation}
f^* =  8\pi \left[ 1 - {0.602033\over N}+O\left( {1\over N^2}\right)
\right],
\label{fstarlnd2}
\end{equation}
should provide more precise estimates. Indeed it is already 
a good approximation for $N=3$ and $N=4$,
where it gives $f^*=20.09$ and $f^*=21.35$ respectively.

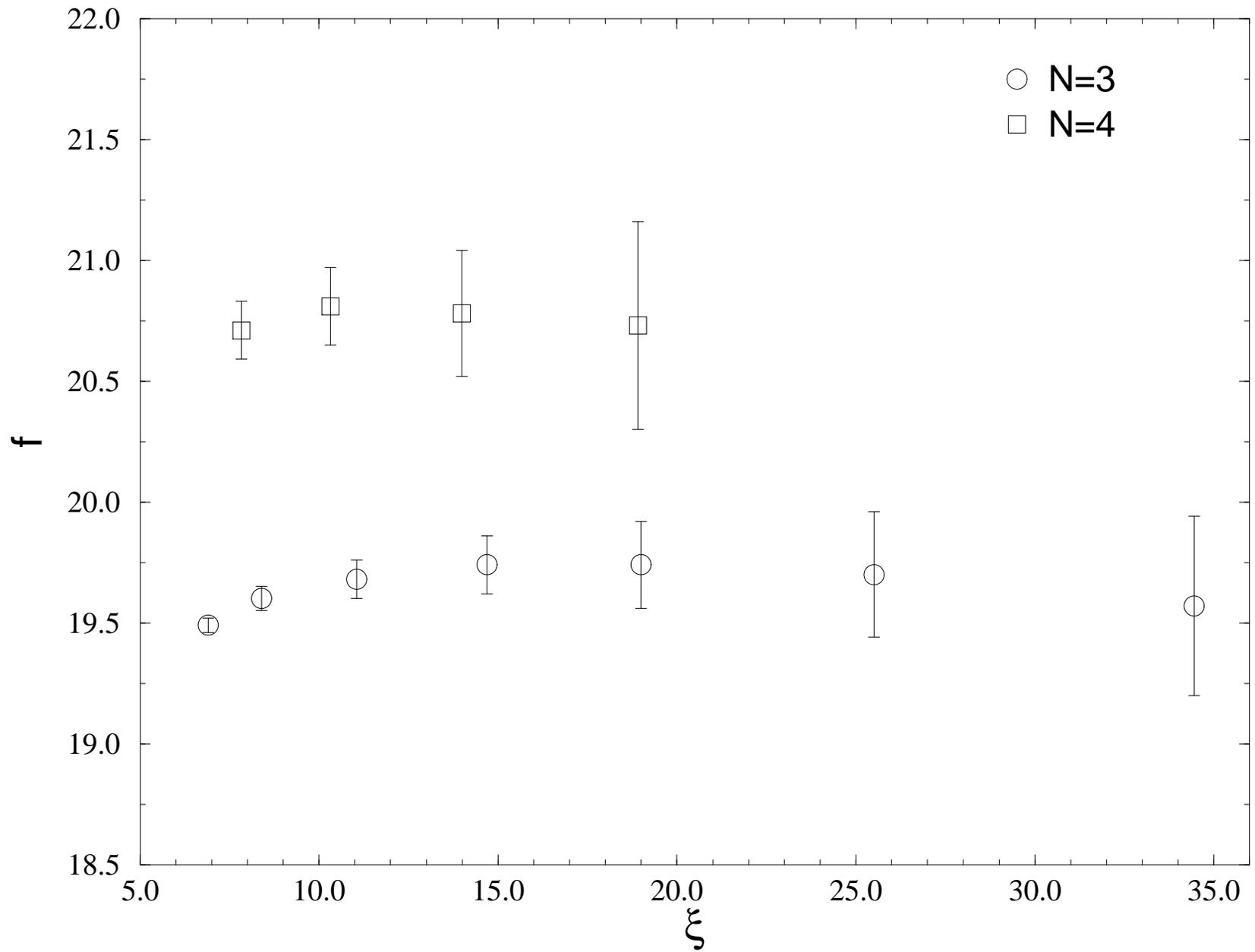
\begin{figure}
\caption{
Two-dimensional O$(N)$ models with $N=3$ and $N=4$:
Plot of $f(\beta)$ vs. $\xi$ as obtained by the analysis
of its strong-coupling series.
}
\label{fig2}
\end{figure}

The reader should notice the small uncertainty of the 
estimates for $N=3,4$  in 
spite of the fact that $\beta_c=+\infty$. This is due to the fact that,
according to field theory, $f(\beta)$ like 
any dimensionless renormalization-group invariant quantity
behaves as
\begin{equation}
f(\beta)-f^*\sim {1\over \xi(\beta)^2},
\label{scalR}
\end{equation}
for sufficiently large $\beta$. 
Hence the corrections to $f^*$ decrease exponentially 
in $\beta$.
This important point was overlooked in Ref.~\cite{B-C-2d}. 
As a consequence of Eq.~(\ref{scalR}) the scaling region, 
where the function $f(\beta)$ 
approximately reaches the asymptotic value,
may begin quite early.
One may obtain good estimates of the
dimensionless renormalization-group invariant quantities 
already at $\xi\gtrsim 10$,
which is still within the reach of the strong-coupling extrapolation, 
at least for not too large values of $N$, say $N=3,4$~\cite{ON-d2-a}.
For instance in Fig. \ref{fig2} we show $f(\beta)$ for $N=3,4$ versus
the correlation length. Scaling is nicely verified 
by our strong-coupling calculations. Our estimates 
of $f^*$ at $N=3,4$ are obtained 
from $f(\beta)$ at $\xi\approx 10$.
The behaviour (\ref{scalR}) explains the success of 
the strong-coupling method when applied to dimensionless
renormalization-group invariant quantities 
(for other examples see Ref.~\cite{ON-d2-letter}). 

We finally
mention the estimates of $f^*$ for $N=2$ and $N=3$ obtained by a Monte Carlo 
simulation together with a finite-size scaling extrapolation~\cite{Kim}.
For $N=2$, fitting the data of Ref.~\cite{Kim} with $\xi \gtrsim 10$ 
to a constant, we get
$f^*=17.7(3)$. For $N=3$ the result is $f^*=19.8(3)$. They are in good 
agreement with the strong-coupling results presented above.

\section{Conclusions}
\label{sec6}

We have studied some issues concerning the fixed-point value
of the zero-momentum four-point renormalized coupling $g$
in O($N$) models. The coupling $g$ plays an important role in 
the field-theoretic perturbative expansion
at fixed dimension, which provides an accurate description
of the symmetric phase. In this approach the value of $g^*$ 
is essential to compute the critical exponents, which are
obtained by evaluating appropriate anomalous dimensions 
(calculated as functions of $g$) at $g^*$.

The first important issue we have discussed is related to the 
presence of confluent singularities at the zero of the
Callan-Symanzik $\beta$-function. In order to understand this
problem we have considered the framework of the $1/N$ expansion,
which provides an analytic and non-perturbative approach.
The analysis of the next-to-leading order of the
$\beta$-function shows the presence of confluent singularities
at its zero, as argued by Nickel~\cite{NickelCS}. 
In generic dimensions $d\neq 4-2/n$ (with $n\in {\N}$),
the leading non-analytic corrections 
are $O(1/N)$ and are related to the exponents
$\Delta_{2,3}/\Delta$  and/or $1+1/\Delta$.
Since they are degenerate at $N=\infty$, one cannot distinguish
them by an $O(1/N)$ calculation.
No term associated
with an exponent $1/\Delta$ is found.
In three dimensions 
one meets a phenomenon analogous to the so-called Abe-Hikami
anomaly~\cite{Abe-Hikami}. One indeed finds that for $d=3$ 
the non-analytic contributions are
$O(1)$ in the $1/N$ expansion, even if the $\beta$-function
appears analytic to leading order. This is essentially
due to a further degeneracy occurring for $d=3$ at $N=\infty$
among $\Delta_{2,3}/\Delta$, $1+1/\Delta$ and
 the analytic correction with exponent two.

In the analysis of the $g$-expansion 
performed by Le Guillou and Zinn-Justin~\cite{LeG-ZJ}
with an additional hypothesis of analyticity, 
the presence of such singularities may cause a slow convergence to
the correct fixed-point value, thus leading to an underestimation
of the real uncertainty~\cite{NickelCS}.  
An accurate check of the $g$-expansion results 
was our major motivation
for the extension of the $\epsilon$-expansion of $g^*$ to $O(\epsilon^4)$,
and for a reanalysis of the strong-coupling expansion 
in the lattice $N$-vector models.

We obtained rather accurate estimates (with an apparent
precision of approximately one per cent, see Table~\ref{summary3d}) 
from the  analysis of the 4th order $\epsilon$-expansion of $g^*$ 
that exploits the known values for  
O($N$) models in lower dimensions.
$g^*$ is indeed expected to be analytic in the 
domain $0<d < 4$, as can be verified in the
large-$N$ expansion to $O(1/N)$.
We plan to extend this analysis to the 
low-magnetization expansion of the effective potential,
which is parametrized by the zero-momentum
$n$-point renormalized couplings. 

The agreement with the $g$-expansion estimates is globally
good. For $N\geq 2$ there is full agreement.
The results for $N=0,1$ are slightly
lower than the estimates given by Le Guillou and Zinn-Justin~\cite{LeG-ZJ},
thus favouring the more general analysis done by Nickel~\cite{Nickelunp}.
This would lead to a small change in the
estimates of the critical exponents, since they 
depend crucially on the value of $g^*$.
For instance, consider the case $N=0$. 
For this model a very precise estimate\footnote{It is worth mentioning
that recently a very precise estimate of $\nu$ 
was obtained by Belohorec and Nickel~\cite{Nickellast}
by a Monte Carlo simulation of the Domb-Joyce model:  $\nu=0.58758(7)$.}
of the exponent $\gamma$ has
been recently obtained by a Monte Carlo simulation: 
$\gamma=1.1575(6)$~\cite{C-C-P}.
On the other hand, 
the analysis of the $g$-expansion, i.e.
the $\beta$-function to $O(g^7)$ and the function
$\gamma(g)$ to $O(g^6)$, of Ref.~\cite{LeG-ZJ} led 
to $\gamma=1.1615(20)$. A more precise 
estimate is reported in
Ref.~\cite{ZJbook}, 
$\gamma=1.1607(12)$: it is  
obtained using the same resummation 
method but one additional order 
in the series of $\gamma(g)$~\cite{Nickelunp}.
Reanalyzing the $O(g^7)$ series of $\gamma(g)$,
using the same method of Ref.~\cite{LeG-ZJ},
the authors of Ref.~\cite{C-C-P} reported the estimate
\begin{equation}
\gamma=1.1616 + 0.11 (\bar{g}^* - 1.421)\pm 0.0004,
\label{CCP}
\end{equation}
where  $\bar{g}^*$ is kept arbitrary.
Our result for $\bar{g}^*$, i.e. $\bar{g}^*\simeq 1.39$,
thus suggests a lower value for $\gamma$,
$\gamma \simeq 1.158$, 
in substantial agreement with the results
of the Monte Carlo simulations and with the analysis of the
$\epsilon$-expansion (see footnote~\ref{expfn}).
Of course more precision and therefore longer series
are necessary to be conclusive.
We also mention that a recent analysis of the
21st order strong-coupling expansion biasing $\Delta$ 
to the known approximate value
has given $\gamma=1.1594(8)$ on the cubic lattice, and
$\gamma=1.1582(8)$ on the b.c.c. lattice~\cite{B-C}
(slightly larger values have been obtained by unbiased analyses).

As a by product of the $\epsilon$-expansion we obtained rather accurate
estimates for the two-dimensional models, which are in good agreement with 
other estimates from lattice approaches (both strong-coupling
expansion and Monte Carlo simulations) and $1/N$-expansion.

Finally we reanalyzed the strong-coupling expansion
of three-dimensional lattice $N$-vector models. In order to get
accurate estimates of $g^*$, we employed an analysis
able to handle the leading confluent singularity. 
For this purpose we used a generalization~\cite{A-M-P}
of the Roskies method~\cite{Roskies}
consisting in a appropriate change of variable.
Final results have an apparent precision of approximately one per cent.
We found good agreement with the field-theoretic estimates.
At $N=0,1$ the results seem to favor the lower
values of $g^*$ of the $\epsilon$-expansion.

\acknowledgments

We gratefully acknowledge useful 
discussions with Paolo Rossi, Lothar Sch\"afer and Alan Sokal.
We thank Michael Fisher for useful comments on the first
draft of this work.

\appendix

\section{Asymptotic expansion of large-$N$ integrals} 
\label{appunosuN}

In this appendix we give a few technical details on the computations 
of Sec.~\ref{sec3}.

The basic problem is the determination of the asymptotic expansion 
for $\widehat{g}\to \widehat{g}^{*}_\infty$ of integrals of the form 
\begin{equation}
I_n(f,\Theta) = 
    \int {d^d u\over (2\pi)^d} 
    {f(u^2)\over \left[1 + \widehat{g}\Pi(u)\right]^n}
\end{equation}
where $f(u^2)$ is a rational function of $u^2$ and 
$\Theta = (\widehat{g}^{*}_\infty - \widehat{g})/\widehat{g}$, for 
$2 < d < 4$. Using the definition 
of $\Pi(u)$, cf. Eq.~(\ref{defPi}), one can rewrite 
\begin{equation}
I_n(f,\Theta) = \left(\widehat{g}^{*}_\infty \over \widehat{g}\right)^n N_d 
     \int_0^\infty u^{d-1} du {f(u^2) \over \left[\Theta + 
\delta(u)\right]^n}
\end{equation}
where $N_d$ is defined in Eq. (\ref{defNd}) and
\begin{equation}
\delta(u) = \left(1 + {u^2\over4}\right)^{d/2 -2} 
   {}_2F_1\left(1 - {d\over2},{1\over2};{3\over2};{u^2\over 4 + u^2}\right).
\label{def-delta}
\end{equation}
We are interested in the large-$u$ behaviour of $\delta(u)$.
Standard identities for the hypergeometric function give
\begin{equation}
\delta(u) = \left({2\over u}\right)^{4-d} c\left[1 + A(u)\right]
\label{deltaasym}
\end{equation}
where $c$ is defined in Eq. (\ref{defc}) and $A(u)$ has the following 
asymptotic expansion
\begin{equation}
A(u) = \sum_{k=1}^\infty {a_k\over u^{2k}} + 
     \left({u\over2}\right)^{4-d} \sum_{k=1}^\infty {b_k\over u^{2k}} ;
\end{equation}
the coefficients $a_k$ and $b_k$ can be easily computed for 
any value of $d$. From Eq. (\ref{deltaasym}) one immediately sees that,
for $d>2$, $\delta(u) \sim u^{d-4}$ for large values of $u$.

For $d<4$ the singularities of $I_n(f)$ are due 
to the large-$u$ domain. Indeed $\delta(u)$ goes 
to zero for $u\to\infty$ and thus one cannot perform
a naive expansion in powers of 
$\Theta$, since for $k$ large enough $I_k(f,0)$ 
diverges.  To compute the asymptotic expansion, let us consider 
\begin{equation}
R(t,u;\Theta) \equiv
{f(t^2 u^2)\over [\Theta + t^{4-d} \delta(t u)]^n} 
  \exp\left[c \left({2\over tu}\right)^{4-d}\right]
\end{equation}
and its expansion for $t\to\infty$ which has the generic form
\begin{equation}
R_{\rm exp}(t,u;\Theta) = \sum_{h,k,p\ge 0} r_{h,k,p} 
      {(tu)^{-2 k - p(4-d)} \over [\Theta + c(2/u)^{4-d}]^{n+h}}.
\end{equation}
Then consider for $2 < d < 4$
\begin{equation}
\int_0^\infty u^{d-1} du\, \left\{ 
    {f(u^2)\over (\Theta + \delta(u))^n} - 
       R_{\rm exp}(1,u;\Theta) 
\exp\left[- c \left({2\over u}\right)^{4-d}\right]
                         \right\}
\label{int-sottratto}
\end{equation}
Because of the exponential factor, the second term 
is integrable for $u\to 0$. 
One can expand the integral (\ref{int-sottratto})
in powers of $\Theta$. Indeed  the subtraction guarantees
that each term is expressed in terms of a convergent 
integral\footnote{The reader 
could rightly be worried by the presence of an infinite series in 
(\ref{int-sottratto}). Indeed all formulae should be intended in 
a formal sense. More precisely the procedure is the following: 
given $K$, to obtain the expansion of $I_n(f)$ to order $\Theta^K$,
one should consider in $R_{\rm exp}(t,u;\Theta)$ only those terms which,
for $\Theta = 0$, decrease for $u\to\infty$ less than or as
$u^{-d-K(4-d)}$. This truncated expansion should be used 
in (\ref{int-sottratto}).}. Thus the integral
(\ref{int-sottratto}) gives rise only to analytic contributions. The 
non-analytic terms can thus be computed from the subtracted term.
Defining 
\begin{equation}
x \equiv  c \left({2\over u}\right)^{4-d} ,
\end{equation}
we see that we must compute integrals of the form
\begin{equation}
J_{n,\alpha} = \int_0^\infty dx\,
   {x^\alpha e^{-x}\over (\Theta + x)^n} 
\end{equation}
with $\alpha\geq 0$.
The computation is now trivial as 
\begin{eqnarray}
 J_{1,\alpha} &=& \Theta^\alpha e^\Theta\Gamma(\alpha+1) 
          \Gamma(-\alpha,\Theta) ,\\
 J_{n,\alpha} &=& {(-1)^{n-1}\over (n-1)!} 
     {d^{n-1}\over d\Theta^{n-1}} J_{1,\alpha} ,
\end{eqnarray}
where $\Gamma(-\alpha,\Theta)$ is the incomplete $\Gamma$-function
\cite{Gradshteyn}.
For $\Theta\to 0$ we have
\begin{eqnarray}
\Gamma(-\alpha,\Theta) = 
   \Gamma(-\alpha) - 
    \Theta^{-\alpha} \sum_{n=0}^\infty {(-\Theta)^n\over n! (n-\alpha)}.
\end{eqnarray}

\section{$\epsilon$-expansion calculation} \label{Appepsilon}

In this Section we report the results of our calculation of 
${g}^*(\epsilon)$ to order $O(\epsilon^4)$,
$\epsilon$ being defined as $\epsilon=4-d$. 

The computation requires the determinations of the massive two-point
and four-point functions for $p^2\rightarrow 0$ to three loops.
The most difficult graphs are those reported in 
Fig.~\ref{fig3}. The calculation is 
straightforward albeit long.
We thus simply report the results:
\begin{eqnarray}
(a) &=& {1\over 3\epsilon^3} - {1\over 6\epsilon^2} + 
     {1\over \epsilon}
        \left(- {1\over12} -{\lambda\over2} + {\pi^2\over 24}\right) - 
     {1\over8} - {\pi^2\over16} + S_1 + {S_2\over 4} + 
     {S_4\over4} - {\zeta(3)\over6},
\\
(b)&=&(a) ,\\
(c) &=& {1\over 6 \epsilon^3} + 
     {1\over \epsilon} 
        \left({1\over 24} - {\lambda\over4} + {\pi^2\over 48}\right) +
     {1\over24} + {11\lambda\over16} + {\pi^2\over24} + 
     {S_1\over2} + {S_2\over8} - {S_4\over4} + {3 S_7\over8} - 
     {5\over24} \zeta(3),
\\
(d) &=& - {5\over 12\epsilon^2} - {5\over 48} - {9\lambda\over32} -
        {5\pi^2\over96} + {5\over2} S_1 + {5\over8} S_2 - {3\over2} S_3 
        - {15\over4} S_5 - {15\over16} S_6 - {3\over16} S_7 
        + {3\over 8} \zeta(3),
\\
(e) &=& {1\over 3\epsilon^3} - {1\over 3\epsilon^2} + 
     {1\over\epsilon}\left({1\over12} + {\pi^2\over 24}\right) 
    - {\pi^2\over24} + S_3 - {2\over3} \zeta(3).
\end{eqnarray}
Each result should be additionally multiplied by $m^{-3\epsilon} N_d^3$
where $N_d$ is defined in Eq. (\ref{defNd}), and $m$ is the mass.
The constants $\lambda$ and $S_i$ are defined by the following integrals:
\begin{eqnarray}
\lambda &\equiv& - \int_0^\infty dt\, {\log t\over t^2 - t + 1}  = \, 
         {1\over3} \psi'\left({1\over3}\right) - {2\pi^2\over9} \approx 
       1.17195, \\
S_1 &\equiv& \int_0^\infty p^3 dp\, \log p\,  d(p)^2\ (L(p) - 2 \log p) \approx
        1.0207, \\
S_2 &\equiv& \int_0^\infty p^3 dp\, d(p)^2\int_0^1 dx\,
 \left[ \log^2 (p^2 x(1-x) + 1) - \log^2(p^2 x(1-x))\right]  \approx 
    - 0.8619, \\
S_3 &\equiv& \int_0^\infty dx \left[ x K_0(x)^2 K_1(x)^2 - 
     {e^{-x/2}\over x} \left(\gamma_E + \log{x\over2}\right)^2\right] \approx 
     0.45077, \\
S_4 &\equiv& \int_0^\infty p^3 dp\, d(p)^2\ (L(p)^2 - 4 \log^2 p) \approx 
     5.9622, \\
S_5 &\equiv& \int_0^\infty p^3 dp\, \log p\ d(p)^3 (L(p) - 2) \approx 
     0.18604, \\
S_6 &\equiv& \int_0^\infty p^3 dp\, d(p)^3\int_0^1 dx\,
        \log^2 (p^2 x(1-x) + 1) \approx 0.21105, \\
S_7 &\equiv& \int_0^\infty p^3 dp\, d(p)^3 (L(p) - 2)^2 \approx 0.19533,
\end{eqnarray}
where $\gamma_E$ is the Euler constant, $\gamma_E \approx 0.577216$,
$K_0(x)$ and $K_1(x)$ are modified Bessel functions, and
\begin{eqnarray}
d(p) &\equiv& {1\over p^2 + 1} , \label{defdp} \\
L(p) &\equiv& \xi \log\left( {\xi + 1 \over \xi - 1}\right) \; ,\\
\xi  &\equiv& \sqrt{1 + {4\over p^2}} .
\end{eqnarray}
The ``Mercedes'' graph $(f)$ requires more sophisticated techniques.
To compute it, we used the method 
of Kotikov \cite{Kotikov}. Let us define
\begin{equation}
I(M,m) = \int {d^dp\over (2\pi)^d} {d^dq\over (2\pi)^d} {d^dr\over (2\pi)^d}
              D(p,M) D(q,M) D(r,M) D(p-q,m) D(p-r,m) D(q-r,m) 
\end{equation}
where $D(p,M)$ is the massive free propagator with mass $M$. 
For $m=0$ this integral was exactly computed in Ref.~\cite{Broadhurst} 
in all dimensions $d$.
The expansion for $\epsilon \to 0$ is given by
\begin{equation}
I(M,0) = M^{-3\epsilon} N^3_d \left({1\over 2\epsilon} \zeta(3) - 
      {\pi^4\over 80} + O(\epsilon)\right) .
\end{equation}
Using the strategy of Ref. \cite{Kotikov} we end up with 
\begin{equation}
I(M,M) = 3 M^{-3\epsilon}
\int_0^{M^2} dm^2 {f(m,M) - f(0,M)\over m^2 (3 M^2 - m^2)}
   + I(M,0) \left(1 + {3\epsilon\over 2} \log{3\over2} + 
        O(\epsilon^2) \right) ,
\end{equation}
where $f(m,M)$ is the following quantity (finite for $\epsilon \to 0$)
\begin{eqnarray}
f(m,M) &=& - M^2 N_d \int_0^\infty q^{d-1} dq 
    \left[ {K(m,M;q^2) J(m,M;q^2)\over q^2 + M^2} \right. 
\nonumber \\
     && \left.
     -{K(M,M;q^2) J(m,m;q^2)\over q^2 + m^2} + 
     {1\over 32\pi^2} {K(M,M,q^2)\over q^2 + m^2}\log{m^2\over M^2}
     \right] .
\end{eqnarray}
The functions $J$ and $K$ can be easily computed for $\epsilon=0$ 
from the integrals
\begin{eqnarray}
K(m,M;q^2) &=& \int {d^dp\over (2\pi)^d} D(p,M)^2 D(p+q,m)  ,
\\
J(m,M;q^2) &=&
- (mM)^{(d-4)/2} (4\pi)^{-d/2} \Gamma\left( 2-\case{d}{2}\right) +
   \int {d^dp\over (2\pi)^d} D(p,M) D(p+q,m).
\end{eqnarray}
A numerical computation gives 
\begin{equation}
I(M,M) = N^3_d M^{-3\epsilon} \left({1\over 2\epsilon} \zeta(3) + H\right)
\end{equation}
with $H = - 0.9825$.

Collecting everything together, we can compute the first four coefficients 
in the expansion (\ref{gepsseries}) of $\bar{g}^*$. Explicitly
\begin{eqnarray}
\bar{g}_0 &=& 1 ,\\
\bar{g}_1 &=& {3(3 N + 14)\over (N + 8)^2}  ,  \\
\bar{g}_2 &=& {1\over (N+8)^4} 
   \left[ -2 N^3 + 58 N^2 + 520 N + 1224 
          -{1\over 3} (13 N + 62) (N+8)^2 \lambda \right.
\nonumber \\
&& \qquad \left. - 12 (5 N + 22) (N+8) \zeta(3) 
    \vphantom{{1\over3}} \right], \\ 
\bar{g}_3 &=& {1\over 8} {1\over (N+8)^6} 
    \left(4 N^5 - 99 N^4 + 5404 N^3 + 57572 N^2 + 225312 N + 341312\right)
\nonumber \\
&& \quad +  {4\over (N+8)^5} 
    \left(23 N^3 - 209 N^2 - 2954 N - 6580\right) \ \zeta(3)
\nonumber \\
&& \quad + {1\over 24} {1\over (N+8)^4} 
  \left[ (11 N^3 - 4338 N^2 - 45312 N - 109472)\lambda + 
     960 (2 N^2 + 55 N + 186) \zeta(5) \right]
\nonumber \\
&& \quad  + {1\over (N+8)^3}
  \left[ - {\pi^4\over 15} (5 N + 22) - {\pi^2\over 3} (19 N + 62) 
         + 16 (N-1) S_3  \right. 
\nonumber \\
&& \quad \quad \left. + 2 (19 N + 62) S_4 - 
        {1\over4} (17 N^2 + 410 N + 1328) S_7 - 8 (5 N + 22) H
   \vphantom{{\pi^4\over 15}}\right]
\nonumber \\
&& \quad + {1\over (N+8)^2} 
    \left[ 24 S_1 + 6 S_2 + 13 (N+2) S_5 + {13\over4} (N+2) S_6\right].
\end{eqnarray}
Numerical values are reported in Sec.~\ref{sec4.1}.

We can use the results for $N\rightarrow\infty$ of Ref.~\cite{ONgr}
to check our expression. For large values of $N$, 
$\bar{g}^{*}$ is given by
\begin{equation}
\bar{g}^{*} = 1 + {1\over N} \left\{ (3-d) 2^{d-1} + 8 -
     \int {d^du\over (2\pi)^d} 
     {\widehat{g}^{*}_\infty\over 1 + \widehat{g}^{*}_\infty \Pi(u)} \, d(u)^2 
 \left[4 d(u) + 9 \left({d\over2} - 1\right) {1\over u^2 + 4} \right] \right\}
\label{eqb27}
\end{equation}
where $\widehat{g}^{*}_\infty$ and $\Pi(u)$ are defined 
in Eqs.~(\ref{defhatgstar}) and (\ref{defPi}) and $d(u)$ in Eq.~(\ref{defdp}).
To obtain the series 
in $\epsilon$, we first expand the
denominator in Eq.~(\ref{eqb27})
in powers of $\widehat{g}^{*}_\infty$ which is of order 
$\epsilon$. The computation is simple. The only integrals 
which require some manipulations are 
\begin{eqnarray}
&& \int {d^du\over (2\pi)^d}\ d(u)^2 \
        {\widehat{g}^{*}_\infty\Pi(u)\over u^2 + 4} = 
   {\epsilon}\left[- {1\over 6} - {\lambda\over 6} + {4\over9}\log2\right]
\nonumber \\
&& \qquad\qquad  +
   \epsilon^2\left[-{1\over6} - {\lambda\over24} + {S_5\over2} + 
         {S_6\over8} + {4\over9}\log 2 - {2\over9} \log^2 2\right] 
     + O(\epsilon^3), 
\\ [2mm]
&& \int {d^du\over (2\pi)^d}\ d(u)^2 \
        {[\widehat{g}^{*}_\infty\Pi(u)]^2\over u^2 + 4} =
    \epsilon^2\left[-{1\over6} - {5\lambda\over48} + {S_7\over8} 
          + {4\over9} \log 2\right] +  O(\epsilon^3).
\end{eqnarray}
The final result is in agreement with our expression.



\begin{table}
\caption{Three-dimensional estimates of $\bar{g}^{*}$ 
from an unconstrained analysis, ``unc", and constrained analyses in
various dimensions. For the analyses which use the
estimates in $d=2$
we report two errors: the first one gives the uncertainty of 
the resummation of the series, the second one expresses the 
change in the estimate when the two-dimensional result varies 
within one error bar.
}
\label{epsilon3d}
\begin{tabular}{cr@{}lr@{}lr@{}lr@{}lr@{}lr@{}l}
\multicolumn{1}{c}{$N$}&
\multicolumn{2}{c}{unc}&
\multicolumn{2}{c}{$d=1$}&
\multicolumn{2}{c}{$d=0,1$}&
\multicolumn{2}{c}{$d=2$}&
\multicolumn{2}{c}{$d=1,2$}&
\multicolumn{2}{c}{$d=0,1,2$}\\
\tableline \hline
0 &  1&.37(9) &  1&.39(3) & & & 
     1&.392(23+1) & 1&.390(16+1) & & \\
 
1 &  1&.37(9) &  1&.41(2) &  1&.39(3) & 
     1&.400(17+0) & 1&.395(16+0) & 1&.397(8+0) \\ 

2 &  1&.36(7) &  1&.39(2) &  1&.41(2) & 
     1&.401(15+2) & 1&.411(12+3) & 1&.413(8+5) \\ 
 
3 &  1&.35(8) &  1&.37(2) &  1&.38(2) & 
     1&.379(11+1) & 1&.386(7+3) & 1&.387(3+4) \\ 

4 &  1&.33(5) &  1&.35(2) &  1&.36(2) & 
     1&.357(11+3) & 1&.364(11+6) & 1&.366(7+8) \\ 

8 &  1&.29(5) &  1&.28(3) &  1&.28(2) & 
     1&.289(15+2) & 1&.295(4+3) & 1&.299(7+5) \\ 

16 & 1&.20(2) &  1&.198(13) & 1&.196(10) & 
     1&.198(8+0)   & 1&.198(4+1)   & 1&.199(1+2) \\ 
 
24 & 1&.154(12) &  1&.150(10) & 1&.148(8) & 
     1&.149(7+0) & 1&.148(5+0) & 1&.148(3+0) \\ 

32 & 1&.123(10) &  1&.121(8) & 1&.119(7) & 
     1&.120(6+0) & 1&.118(4+0) & 1&.118(3+0) \\ 

48 & 1&.087(8) &  1&.086(5) & 1&.085(5) & 
     1&.085(4+0) & 1&.084(3+0) & 1&.084(2+0) \\ 
\end{tabular}
\end{table}

\begin{table}
\caption{Two-dimensional estimates of $\bar{g}^{*}$ 
obtained from analyses constrained at $d=1$ and at $d=0,1$.
}
\label{epsilon2d}
\begin{tabular}{cr@{}lr@{}l}
\multicolumn{1}{c}{$N$}&
\multicolumn{2}{c}{$d=1$}&
\multicolumn{2}{c}{$d=0,1$}\\
\tableline \hline
0 &  1&.69(7) & & \\ 
1 &  1&.79(5) &  1&.75(5) \\ 
2 &  1&.75(5) &  1&.79(3) \\ 
3 &  1&.68(6) &  1&.72(2) \\ 
4 &  1&.61(5) &  1&.64(2) \\ 
8 &  1&.45(5) &  1&.45(2) \\ 
16 &  1&.28(3) &  1&.28(1) \\ 
24 &  1&.21(2) &  1&.20(1) \\
32 &  1&.16(2) &  1&.16(1) \\ 
48 &  1&.11(1) &  1&.11(1) \\ 
\end{tabular}
\end{table}

\begin{table}
\caption{
Estimates of $f^*$ obtained from the analysis
of the available strong-coupling series with (GRT) and without
(ST) the generalized Roskies transform (cf. Eq.~(\ref{cRT})). 
We also report the values of $\beta_c$ and $\Delta$ used in our
GRT analyses.
The apparent uncertainty  in the estimate obtained by
employing the GRT is
expressed as a sum of two numbers: the first number comes 
from the analysis at $b=\Delta$,
the second one is due to the uncertainty on $\Delta$, and it
is obtained by varying $b$. 
For $N=0,1,2$, the estimate of $f^*$ is 
increasing with increasing 
$b$ in the GRT.
\label{SCresults}}
\begin{tabular}{ccr@{}lr@{}lr@{}lr@{}l}
\multicolumn{1}{c}{$N$}&
\multicolumn{1}{c}{lattice}&
\multicolumn{2}{c}{$\beta_c$}&
\multicolumn{2}{c}{$\Delta$}&
\multicolumn{2}{c}{GRT}&
\multicolumn{2}{c}{ST}\\
\tableline \hline
0&cubic & 0&.213492(1)\cite{Grassberger}& 0&.50(5)& 17&.50(19+6) & 19&.1(4) \\

1& cubic & 0&.2216544(6)\cite{Blote}     & 0&.50(5)& 23&.64(14+10) & 24&.9(3)\\

1& b.c.c.& 0&.157373(2)~\cite{B-C}  & 0&.50(5)& 23&.53(9+10)& 24&.9(2) \\

1& f.c.c.& 0&.102062(5)~\cite{Z-L-F}     & 0&.50(5)& 23&.53(18+10)& 24&.6(4) \\

1&diamond& 0&.36969(10)~\cite{d3paper} & 0&.50(5)& 23&.47(24+12) & 25&.1(4) \\

2&cubic  & 0&.22710(1)\cite{Hasenbush}   & 0&.52(2)& 28&.45(17+5) & 29&.4(3) \\

3&cubic  & 0&.231012(12)\cite{Chen-Holm} & 0&.55(2)& 32&.24(21+5) & 32&.39(2)\\

4&cubic  & 0&.23398(2)~\cite{B-C}        & 0&.57(2)& 35&.10(30+10)& 34&.7(1) \\

8&cubic  & 0&.24084(3)~\cite{B-C}        & 0&.66(2)& 41&.50(20+10)& 40&.3(2) \\

16&cubic & 0&.24587(6)~\cite{B-C}        & 0&.77(2)& 45&.81(10+10)& 44&.7(1) \\

24&cubic & 0&.24795(3)                   & 0&.83(2)& 47&.28(6+10) & 46&.4(1) \\

32&cubic & 0&.24907(2)                   & 0&.88(2)& 47&.96(4+10) & 47&.37(10)\\
48&cubic & 0&.25023(2)                   & 0&.93(1)& 48&.66(4+4)  & 48&.32(10)\\
\end{tabular}
\end{table}

\begin{table}
\caption{
Summary of the three-dimensional estimates of $\bar{g}^*$.
In Refs.~\protect\cite{Nickelunp,Antonenko,Sokolov} the results
were reported without errors.
\label{summary3d}}
\begin{tabular}{cr@{}lcr@{}lr@{}l}
\multicolumn{1}{c}{$N$}&
\multicolumn{2}{c}{$\epsilon$-exp.}&
\multicolumn{1}{c}{$g$-exp.}&
\multicolumn{2}{c}{H.T.}&
\multicolumn{2}{c}{$1/N$-exp.}\\
\tableline \hline
0 & 1&.390(17) &
1.421(8)~\cite{LeG-ZJ,ZJbook},$\;$1.39~\cite{Nickelunp}& 
1&.393(20) &&  \\

1& 1&.397(8) &
1.414(3)~\cite{G-Z},$\;$
1.40~\cite{Nickelunp} &1&.406(9) && \\

2 & 1&.413(13) & 
1.406(4)~\cite{LeG-ZJ,ZJbook},$\;$
1.40~\cite{Nickelunp}&
1&.415(11) && \\

3 & 1&.387(7) & 
1.391(4)~\cite{LeG-ZJ,ZJbook},$\;$
1.39~\cite{Nickelunp}&
1&.411(12) && \\

4 & 1&.366(15) &
1.374~\cite{Antonenko,Sokolov} & 
1&.396(16) && \\

8 & 1&.295(7) &
1.304~\cite{Antonenko,Sokolov} &  1&.321(10) && \\

16 & 1&.199(3) &
1.208~\cite{Antonenko,Sokolov}&  1&.215(5) & 1&.204 \\

24 & 1&.148(3) &
1.154~\cite{Antonenko,Sokolov} & 
1&.158(4) & 1&.151 \\

32 & 1&.118(3) &
1.122~\cite{Antonenko,Sokolov} & 
1&.122(3) & 1&.1196 \\

48 & 1&.084(2) & & 1&.084(2) & 1&.0839 \\
\end{tabular}
\end{table}

\begin{table}
\caption{Summary of the two-dimensional estimates of $\bar{g}^*$.
\label{summary2d}}
\begin{tabular}{cr@{}lr@{}lr@{}lr@{}lr@{}l}
\multicolumn{1}{c}{$N$}&
\multicolumn{2}{c}{$\epsilon$-exp.}&
\multicolumn{2}{c}{H.T.}&
\multicolumn{2}{c}{$1/N$-exp.}&
\multicolumn{2}{c}{$g$-exp.}&
\multicolumn{2}{c}{M.C.}\\
\tableline \hline
0 & 1&.69(7) & 1&.679(3) & && && & \\
1 & 1&.75(5) & 1&.7540(2)& && 1&.85(10)~\cite{LeG-ZJ} & 1&.71(12)~\cite{K-P} \\
2 & 1&.79(3) & 1&.810(10)& && & & 1&.76(3)~\cite{Kim} \\
3 & 1&.72(2) & 1&.724(9) & 1&.758 & 1&.749(16)~\cite{Falcionietal} & 
1&.73(3)~\cite{Kim} \\
4 & 1&.64(2) & 1&.655(16)& 1&.698 & & & &\\
8 & 1&.45(2) &  && 1&.479 & & & &\\
16& 1&.28(1) &  && 1&.283 & & & &\\
24& 1&.20(1) &  && 1&.200 & & & &\\
32& 1&.16(1) &  && 1&.154 & & & &\\
48& 1&.11(1) &  && 1&.106 & & & &\\
\end{tabular}
\end{table}

\begin{figure}
\caption{
Three-loop Feynman graphs contributing to the
four-point correlation function.
}
\label{fig3}
\end{figure}

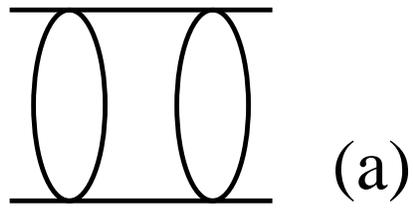
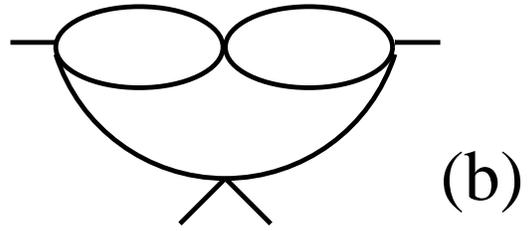
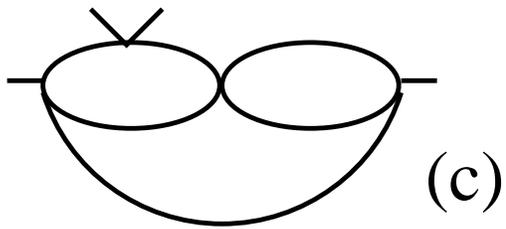
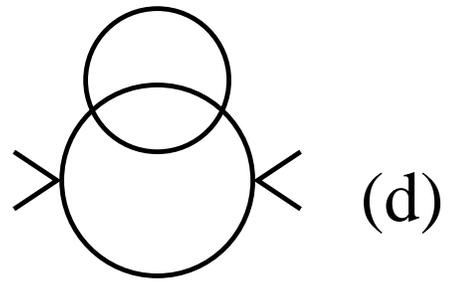
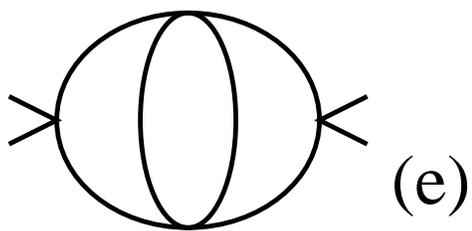
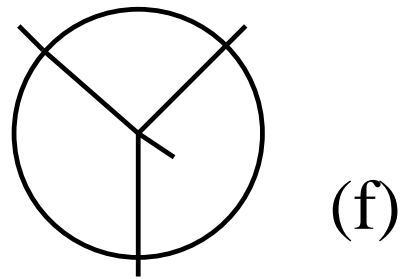
\end{document}